\documentclass[10pt, journal]{IEEEtran}
\IEEEoverridecommandlockouts
\usepackage[noadjust]{cite}

\usepackage{amsmath}
\usepackage{graphicx}
\usepackage{tabularx}
\usepackage{textcomp}
\usepackage{xcolor}
\usepackage{balance}
\usepackage{comment}
\def\BibTeX{{\rm B\kern-.05em{\sc i\kern-.025em b}\kern-.08em
    T\kern-.1667em\lower.7ex\hbox{E}\kern-.125emX}}
\usepackage[english]{babel}
\usepackage[utf8]{inputenc}
\usepackage{algorithm}
\usepackage[noend]{algpseudocode}
\usepackage{pgfplots}
\usepackage{subcaption}
\usepackage{multirow}
\usepackage{adjustbox}

\usepackage{booktabs}
\usepackage{graphicx}

\pgfplotsset{width=8.5cm,compat=1.9}

\include{pythonlisting}

\begin{document}

\title{Defending Hardware-based Malware Detectors against Adversarial Attacks\\}

\author{\IEEEauthorblockN{Abraham Peedikayil Kuruvila, \textit{Student Member, IEEE}, Shamik Kundu, \textit{Student Member, IEEE},\\
Kanad Basu, \textit{Member, IEEE} }


\thanks {A. P. Kuruvila (e-mail: apk190000@utdallas.edu), S. Kundu and K. Basu are with the Department of Electrical and Computer Engineering, University of Texas at Dallas, Richardson, TX, 75080.}

}

\maketitle

\begin{abstract}
In the era of Internet of Things (IoT), Malware has been proliferating exponentially over the past decade. Traditional anti-virus software are ineffective against modern complex Malware. In order to address this challenge, researchers have proposed Hardware-assisted Malware Detection (HMD) using Hardware Performance Counters (HPCs). The HPCs are used to train a set of Machine learning (ML) classifiers, which in turn, are used to distinguish benign programs from Malware. Recently, adversarial attacks have been designed by introducing perturbations in the HPC traces using an adversarial sample predictor to misclassify a program for specific HPCs. These attacks are designed with the basic assumption that the attacker is aware of the HPCs being used to detect Malware. Since modern processors consist of hundreds of HPCs, restricting to only a few of them for Malware detection aids the attacker. In this paper, we propose a Moving target defense (MTD) for this adversarial attack by designing multiple ML classifiers trained on different sets of HPCs.  The MTD randomly selects a classifier; thus, confusing the attacker about the HPCs or the number of classifiers applied. We have developed an analytical model which proves that the probability of an attacker to guess the perfect HPC-classifier combination for MTD is extremely low (in the range of $10^{-1864}$ for a system with 20 HPCs). Our experimental results prove that the proposed defense is able to improve the classification accuracy of HPC traces that have been modified through an adversarial sample generator by up to 31.5\%, for a near perfect (99.4\%) restoration of the original accuracy.
\\
\end{abstract}


\section{Introduction}
\label{sec:intro}
Malicious software, informally known as Malware, include Trojans, worms, spyware, adware, and computer viruses, etc. They intentionally try to harm a system or leak sensitive private information significant to the user. Traditional software-based Malware detection techniques were used in the form of Anti-Virus Software (AVS) to identify whether a program is benign or malicious. Many AVS function by simply running the program in a virtual machine and recording which Application Programming Interfaces (APIs) are used \cite{lou2010application}. Behavior flag values are set by monitoring the functioning of the program. The AVS then classifies the program as either Malware or benign by trying to match the API usage with the behaviors recorded with known data. As computing systems become more complex, attackers have become conscious on the type of attacks they launch \cite{demme2013feasibility}. In response, commercial AVS had to bolster their tools in order to enhance the system security. This has caused AVS to induce significant performance overhead to accommodate the large computational bandwidth \cite{ozsoy2016hardware}. Moreover, AVS have been seen to be ineffective against modern polymorphic and metamorphic Malware \cite{demme2013feasibility}. 

\begin{figure}
\begin{subfigure}{0.5\textwidth}
  \includegraphics[width=0.975\linewidth]{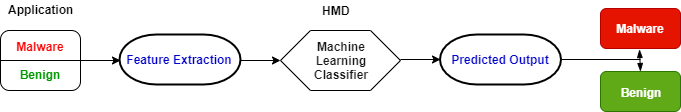}
  \caption{HMD based on HPC Traces.}
  \label{fig:intro1}
\end{subfigure}

\begin{subfigure}{0.5\textwidth}
  \includegraphics[width=0.975\linewidth]{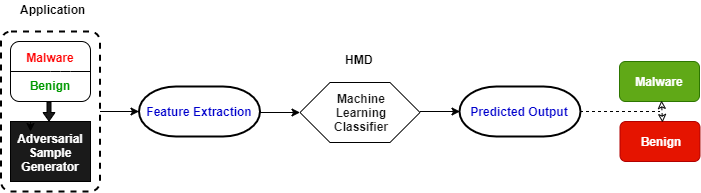} \caption{Adversarial Attack to alter HPC Traces.}
  \label{fig:intro2}
\end{subfigure}
\caption{Process Flow Diagram of Malware Detection.}
\label{fig:intro3}
\end{figure}

In order to address these challenges, researchers are exploring alternatives to AVS for Malware detection. A promising approach in this direction is the application of Hardware-assisted Malware Detection (HMD) \cite{ozsoy2016hardware}. HMDs use dedicated hardware features to distinguish between Malware and benign programs. Hardware features are more difficult to jeopardize compared to their software equivalents. Hardware Performance Counters (HPCs) have emerged as a promising candidate for Malware detection. HPCs are special registers built into all modern processors that count significant low-level micro-architectural features like branch misses, cache misses, and CPU cycles, etc. They are beneficial for conducting power analysis and tuning the performance of a computing system. 
\textcolor{black}{These HPC values, harvested by executing an extensive set of Malware and benign applications, are used to train the ML classifier in a Malware detector, after which it is released to the users. Figure~\ref{fig:intro1} outlines the operation of an already trained Malware detector in post-release phase. In this post-release inference phase; first, the trained HMD executes feature extraction on an unknown input vector, following which it predicts the application to be benign or malicious with the aid of its inbuilt ML classifier, as shown in Figure~\ref{fig:intro1}.} 

Recently, adversarial attacks are being developed on HPC-based Malware detectors. 
\textcolor{black}{In one such attack, as discussed in \cite{dinakarrao2019adversarial}, perturbations are introduced into the HPC traces with an aim to subdue the HMD. It is assumed that the HPCs characterized by the developer in the HMD is private information, and the adversary is obscured from the knowledge and authority to modify the counter values in the HPC trace during execution time. Hence, as stated in \cite{dinakarrao2019adversarial}, the attacker considers the victim's defense system to be a black box and then reverse engineers the HMD so as to replicate the behavior of the security system. This reverse engineering process furnishes an optimal machine learning classifier that utilizes the embedded feature set of harvested HPCs from the application data. The attacker then employs an adversarial sample predictor to predict the number of HPCs to be generated to impel the Malware detector into misclassifying an application. Following this, an adversarial sample generator is developed to execute in conjunction with the target application, which generates the preset HPC values acquired from the adversarial sample predictor. These fabricated values engendered from the adversarial attack result in misclassification, consequently resulting in an eroding of trust in the system. The attack has been visualized in Figure~\ref{fig:intro2}, where the ML classifier, similar to Figure~\ref{fig:intro1}, has been pre-trained and deployed in the post-release inference phase. With the information obtained from reverse engineering and sample predictor, the adversarial sample generator, in the inference phase, fabricates the preset HPC values while being executed in conjunction with the victim application in the same thread of the processor, as depicted in Figure 1b. Since this sample generator code has no malicious features, it will not be detectable by the deployed HMD, and thus the adversarial sample generator stealthily accomplishes the intended attack on the Malware detector in the post-release inference phase. } 


\textcolor{black}{
This paper counters the adversarial attack by establishing a Moving Target Defense (MTD) that utilizes various ML classifiers deployed dynamically. The attack surface available to the attacker is continuously altered, by continually modifying the following parameters employed in Malware detection -- $(1)$ the count of HPCs harvested from the processor, $(2)$ the specific set of HPCs to be provided as input to the HMD, and $(3)$ precisely, which classifier of the HMD is utilising which HPC combinations to distinguish between benign and malicious applications.} As a result of this consistent parameter variation, estimating those on the part of the adversary is practically improbable, thereby jeopardising the planned attack by exponentially increasing the attack complexity, as explained in detail in Section \ref{sec:MTD:CCA}.

\textcolor{black}{The key contributions of this paper are as follows:
\begin{itemize}
    \item This work proposes a Moving Target Defense (MTD) model that dynamically changes the attack surface by modifying the count, the set of HPCs utilised, and the distinct classifier deployed for defending against adversarial attacks on Malware detection, with tolerance to negligible variation in classification accuracy (0$\sim$3\%).
    \item To illustrate the robustness of our proposed defense mechanism, the paper demonstrates a thorough statistical analysis with multivariate HPC and classifier combinations, thereby yielding an infinitesimal probability for reverse engineering the system (as low as $10^{-1864}$ for a processor with 20 HPCs).
    \item The proposed MTD algorithm is evaluated against adversarial attacks on HMDs detecting real life Malware. The experimental results obtained by varying the classifiers based on their algorithms, their performance statistics and the specific set of HPCs utilised by a distinct classifier corroborate our hypothesis of a robust defense strategy.
    \item The performance of the MTD algorithm when evaluated against strengthened adversarial attack variants proves to be resilient, thereby furnishing immense enhancement in security to the processor in lieu of minimal area overhead.
    \item We have synthesized and measured the hardware overhead for the proposed MTD algorithm.
\end{itemize}}

The paper is organized as follows. Section~\ref{sec:background} outlines the adversarial attack on Malware detectors that this paper is defending against. Section~\ref{sec:rel} describes related work on the concepts of MTD as well as HPC-based Malware detection. The proposed MTD-based approach is presented in Section~\ref{sec:MTD}. Section~\ref{sec:Experiment} presents the experimental results. Section~\ref{sec:Discussion} discusses the impact of the proposed defense. Finally, Section~\ref{sec:Conclusion} concludes the paper.

\section{Background}
\label{sec:background}
\subsection{Hardware Performance Counters in Malware Detection}
\label{sec:background-HPCinML}
HPCs are special purpose registers which keep count of crucial low-level architectural events. The HPC values obtained from executing a program can be employed for software optimization and performance tuning of the cyber-physical system. The number of accessible HPCs varies depending on the processor. Prior research in HPC-based HMD has proven that HPCs can be used in conjunction with ML classifiers to efficiently detect malicious applications \cite{wang2015reusing, demme2013feasibility}.

\subsubsection{Hardware Performance Counter Collection}
\label{sec:backgroundHPCcollection}
This subsection explains how HPC data is collected for each application, in order to design the HMD. Linux OS provides a package called \textit{linux-tools-common} which gives access to the {\it perf} command. {\it Perf} is a performance analyzing tool that can be accessed from the command line of the terminal to obtain information on various hardware and software events related to a program. The command {\it perf list} returns a list of all the hardware and software events that \textit{perf} can monitor. 
Executing the command {\it perf stat} with proper parameters for constraints alongside the executable of the application will provide the specified HPC values for that program.
\textcolor{black}{The number of HPCs that can be monitored simultaneously in a processor is limited, typically 4-6, depending on the processor architecture \cite{basu2019theoretical, demme2013feasibility, wang2015reusing}. }
In this paper, we have retrieved four HPC measures per \textit{perf} command from the target processor.

\subsubsection{Machine Learning Classifiers}
\label{sec:backgroundMLclassifiers}
In this section, a brief overview of the conventional ML classifiers is provided, which are deployed in the HMD \cite{dinakarrao2019adversarial,demme2013feasibility}. 
Other ML classifiers could also be used, albeit, with a difference in classification accuracy.

A \textbf{Decision Tree} is an algorithm that builds a classification model in a tree like structure. This ML classification model uses a mutually exclusive if-then rule set, where the rules are learned sequentially from the training data set one at a time \cite{asiri_2018}. The structure of the tree is a top-down recursive  approach where the most dominant attributes are at the top. Decision Trees can easily be over-fitted, so pruning is required to trim unnecessary branches in the tree.

A \textbf{Neural Network} consists of a set of neurons and layers that are interconnected such that each connection has some sort of weight linked with it. When training the Neural Network, the network will learn and adjust the weights in order to be capable of predicting and classifying the correct label for the input tuples. In a Neural Network, there can be many layers in the model, and the number of hidden layers in the model will depend on how complex the model has to be. Increasing the number of hidden layers in the model will consequently increase the time it takes to train the model. However, Neural Networks have demonstrated good performance in practical applications as they have a leniency towards data with noise.

\subsubsection{Measurement metrics}
\label{sec:background-metrics}
Three different measurement metrics are used to evaluate  the performance of the ML classifiers, namely: {\it  accuracy, precision, and recall}.  In this section, we will furnish the formal definition for these metrics.

\textbf{Accuracy} is the ratio of the number of predictions that are classified correctly to the total number of predictions, which can be represented as: \\
\begin{equation}
\label{equ1}
Accuracy =  \frac{No.\  of\ (TP\ +\ TN)}{ No.\  of\ (F  P\ +\ F N\ +\ T P\ +\ T N)\ }
\end{equation}

where, {\it TP} stands for True Positive, {\it TN} stands for True Negative, {\it FP} stands for False Positive, and {\it FN} stands for False Negative in Equation~\ref{equ1}. For a ML classifier that tries to classify whether an application is a Malware (positive class) or benign (negative class), a True Positive would be a Malware that is correctly classified malicious while a True Negative is an application that is correctly labeled benign. Furthermore, a False Positive would be a Malware that is incorrectly labeled benign, while a False Negative would be a benign application that is incorrectly labeled Malware. 

\textbf{Precision} represents the proportion of positive classifications that were correct.  \\
\begin{equation}
\label{equ2}
Precision =  \frac{No.\  of\  T  P}{No.\ of\ F  P\ + No.\ of\ T P\ }
\end{equation}
As an example, a ML model used in a HMD that has a TP score of 10 and a FP score of 10 indicates that the precision of the model is only 50\%.

\textbf{Recall} is the proportion of actual positives that were correct.  In other words, recall is a measure of the fraction of Malware samples that were classified correctly.\\
\begin{equation}
\label{equ3}
Recall =  \frac{No.\  of\  T  P}{No.\ of\ F  N\ + No.\ of\ T P\ }
\end{equation}

As an example, the ML model has TP score of 10 and FN score of 90, signifying that the model only correctly identifies 10\% of the Malware in the dataset. Hence, its recall is 10\%.

\subsection{Adversarial Attack on Hardware Performance Counters}
\label{sec:background-adversarialattack}
While ML classifiers are robust to noise, it has been shown that the output can be modified by introducing perturbations to the input~\cite{goodfellow2014explaining}. These perturbations are denoted as adversarial samples. Recently, an adversarial attack in HPC-based HMDs was proposed and implemented by \cite{dinakarrao2019adversarial}. The authors modified the HPC values remotely by introducing additional instructions to deceive the HMD into misclassifying the applications. This adversarial attack consists of three parts, as follows:

\subsubsection{Reverse Engineering HMD}
\label{sec:background-adversarialattack-reverseHMD}
Before the attack can be fabricated, the attacker needs to reverse engineer the ML classifier used in the HMD.  First, a large data set of benign and Malware programs are tested on the detector. The HMD is used as a black box. The responses from the HMD are recorded and are used as labels to train various ML models. The ML models are then tested, and their output is compared with the HMD's output, from which the adversary is able to deduce if the newly trained ML models are accurate enough to be considered a reverse engineered HMD. The ML classifier with the best accuracy is selected to perform as a reverse-engineered version of the HMD.
\subsubsection{Perturbations in HPCs}
\label{sec:background-adversarialattack-perturbationsHPC}
To compute the number of additional perturbations needed to generate an adversarial attack in the HPC trace, a gradient loss approach, similar to the fast-gradient sign method, is utilized \cite{dinakarrao2019adversarial}.  This approach provides low overhead in terms of computational complexity. For a ML model, a hyper parameter $\alpha$ is assumed. The input and output of the ML models are represented as \textit{b} and \textit{c}, respectively. Thus, the cost function of training the model is $C(\alpha,b,c)$\cite{dinakarrao2019adversarial}. Necessary perturbations to deceive the HMD are obtained by the cost function gradient loss of this model, shown in Equation~\ref{equ4}.

\begin{equation}
\label{equ4}
 Perturbations = b + \epsilon sign (\nabla_b C(\alpha,b,c))        
\end{equation}

In Equation~\ref{equ4}, $\nabla_b$ is the gradient of the cost function with respect to input $b$ while, $\epsilon$ is a variable in range $0$ to $1$. This is performed to keep the variation of input $b$ undetectable. A lower boundary constraint is instantiated on the model, since HPC values cannot be negative.  

\subsubsection{HPC Modification}
\label{sec:background-adversarialattack-HPCmodification}
Adversarial attacks on HMD focus on targeting specific HPC values. This adversarial attack \cite{dinakarrao2019adversarial}  focuses on manipulating branch-misses and LLC-load-misses, which in turn alters the counter values for instructions and branch-instructions. There is no direct access to modifying the HPC values in a program. Instead, the adversarial attack employs a benign program that is wrapped around an application to be misclassified to generate the necessary HPC values needed in the trace. The program to be camouflaged is executed with the adversarial sample generator concurrently, on the same thread and core of the processor.

\begin{algorithm}
\label{sec:algorithm1}
\vspace{1mm}
\textbf{Input}: Application ‘App()’\\
\textbf{Output:} Adversarial micro-architectural events\\
\vspace{-4mm}
\caption{Branch-miss Generation Psuedo-Code}
\begin{algorithmic}[1]
\State \textbf{\#define} int $a,b,c,d,e$
\State $a<b<c<d<e$
\While{$a < Num\_To\_Get\_HPCs$}
\State \textbf{if}  $a>b$ \hspace{5mm} \{----do nothing----\}
\State \textbf{if}  $a>c$ \hspace{5mm} \{----do nothing----\}
\State \textbf{if}  $a>d$ \hspace{5mm} \{----do nothing----\}
\State \textbf{if}  $a>e$ \hspace{5mm} \{----do nothing----\}
\EndWhile
\State \textbf{end}
\end{algorithmic}
\end{algorithm}

In order to generate branch-misses in the HPC trace, Algorithm 1 initializes a few variables such that the first variable is the smallest in value. Dummy {\it if} statements that compare the value of the smallest variable to the value of the other variables are executed. It is already known that the {\it if} statements will fail. Hence, the failure of these {\it if} statements will produce the necessary branch-misses to append to the concerned HPC trace. The number of additional branch-misses to be introduced is computed using the approach described in Section~\ref{sec:background-adversarialattack-perturbationsHPC}. Similarly, introduction of LLC-load-misses into the HPC trace can be accomplished by loading an array of a fixed size. The values in the array are all flushed. The values are then reloaded, and as a result, the LLC-load-misses increase in the HPC trace. After the HPC sample predictor has predicted the number of HPCs needed to be appended to the current HPC trace, and the adversarial HPC generator has been executed concurrently with the applications, the accuracy of the HMD classifier is observed. The modified HPC traces are tested using the reversed engineered HMD to verify the reduction in detection accuracy.

\subsection{Adversarial Attack Results}
\label{sec:background-adattackresults}
\begin{figure}[h!]
\centering
  \includegraphics[width=0.85\linewidth]{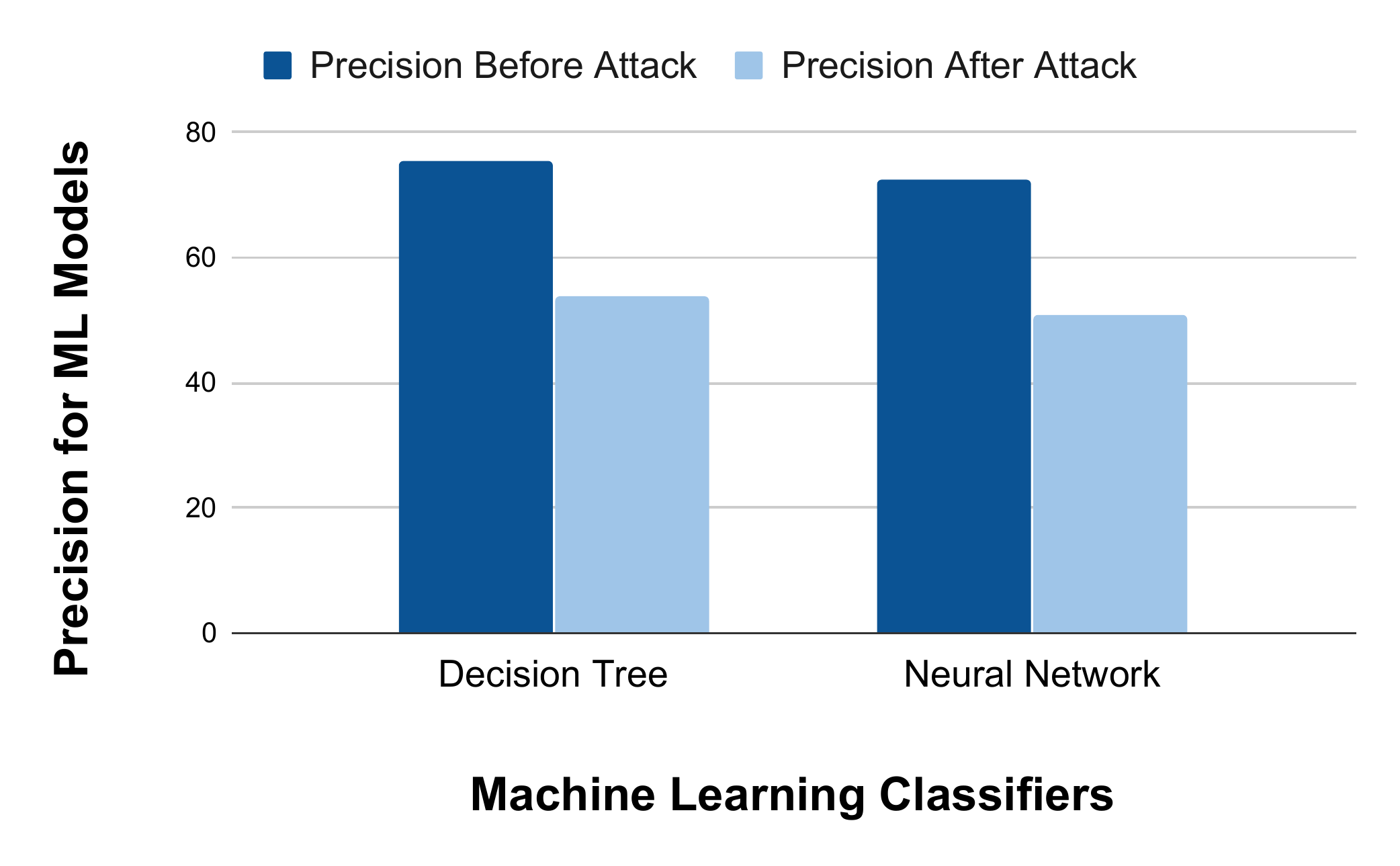}
  \caption{Precision before and after Adversarial Attack.}
  \label{fig:classpre}
\end{figure}

\begin{figure}[h!]
\centering
  \includegraphics[width=0.85\linewidth]{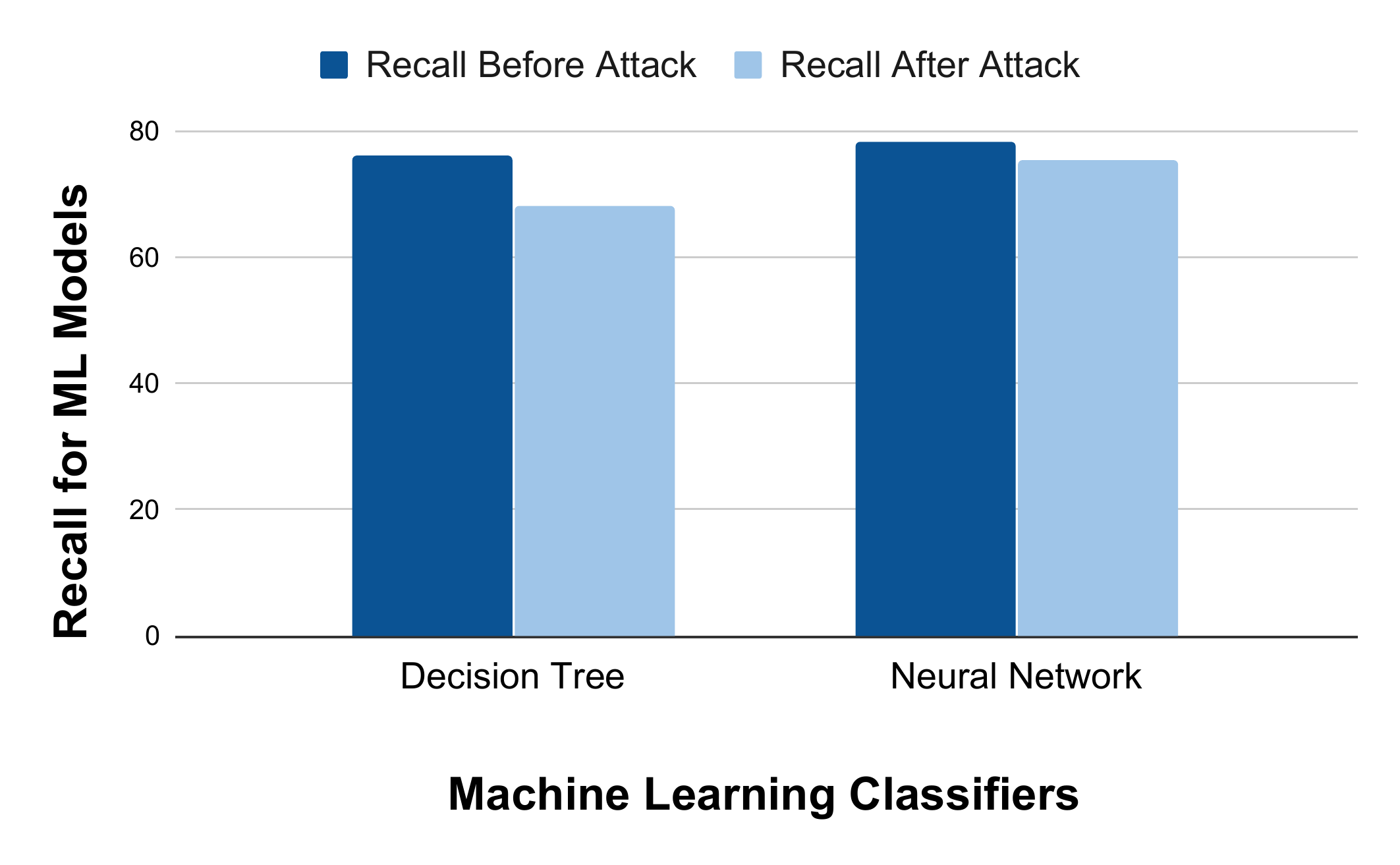}
  \caption{Recall before and after Adversarial Attack.}
  \label{fig:classrec}
\end{figure}

\begin{figure}[h!]
\centering
  \includegraphics[width=0.85\linewidth]{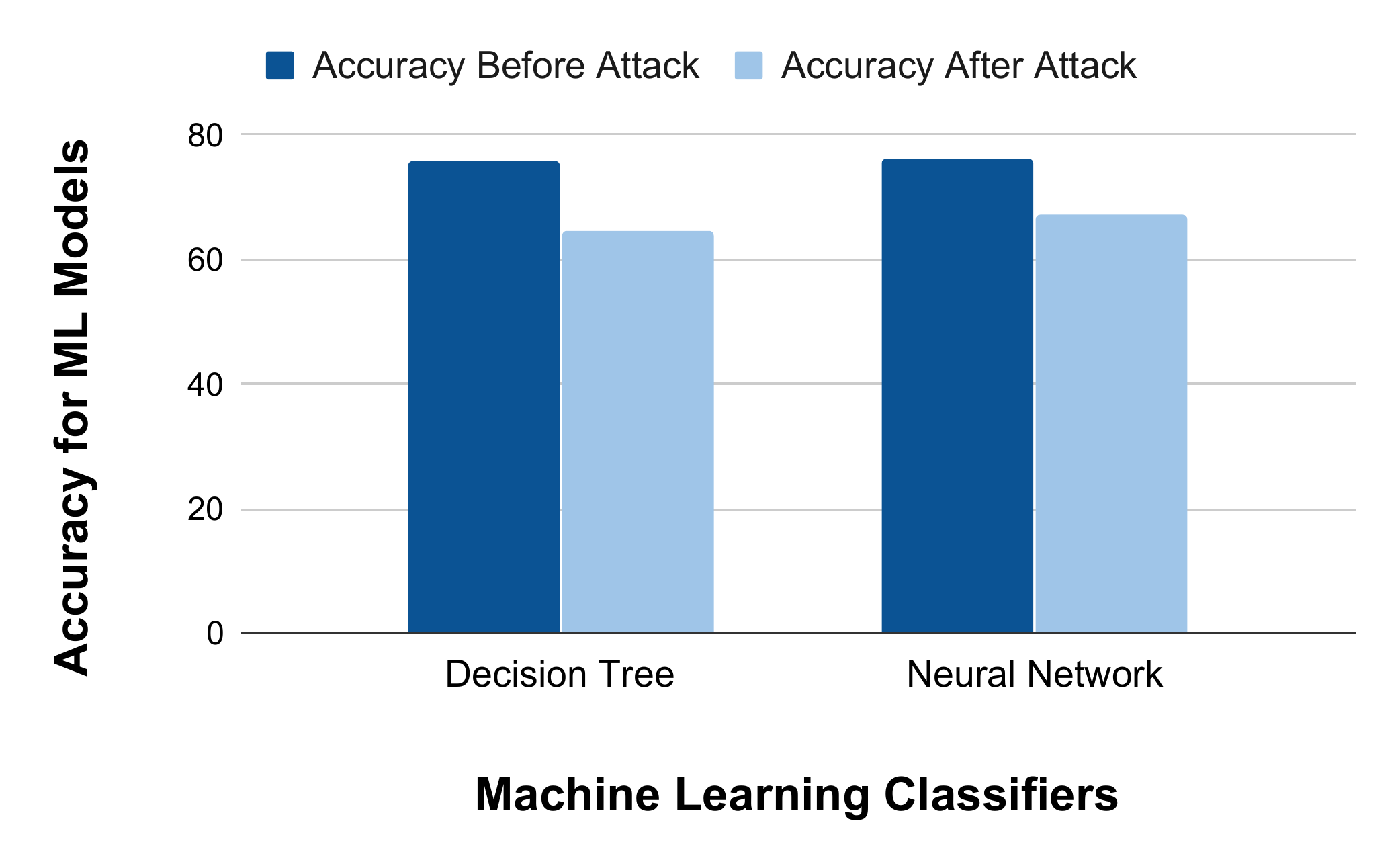}
  \caption{Accuracy before and after Adversarial Attack.}
  \label{fig:classacc}
\end{figure}

In replicating the attack, two ML classifiers, Decision Tree and Neural Network, were trained on the HPCs harvested by executing \textcolor{black}{ 300 Malware and 300 benign applications. From the set of available HPCs, we utilized} HPCs branch-misses, LLC-load-misses, instructions, and branch-instructions, similar to \cite{dinakarrao2019adversarial}. \textcolor{black}{50 new Malware, that were ran with the Adversarial Sample Generator, are then tested alongside 50 new benign programs. Figure~\ref{fig:classpre}, Figure~\ref{fig:classrec}, and Figure~\ref{fig:classacc} show the results for the precision, recall, and accuracy, respectively, for both classifiers. Before the attack, Decision Tree classifier had a precision of 75.65\%, recall of 76.04\%, and an accuracy of 75.8\%, while the Neural Network classifier had a precision of 72.43\%, recall of 78.48\%, and an accuracy of 76.35\%. After the attack, all the performance metrics were affected. However, we focus on the impact of precision for both the classifiers. Since precision furnishes a measure of the True Positive rate, that is, the number of Malware being successfully detected, a reduced precision metric makes way for the Malware to stealthily evade detection by the HMD, thereby posing a threat to the security of the system, unlike their benign counterparts. As a result, the degradation of precision becomes a major concern over the degradation of accuracy, which encompasses the successful detection of both benign and Malware applications. After the attack, the Decision Tree had a precision of 53.96\%, and the Neural Network had a precision of 50.96\%, which is almost similar to a random guess. The Decision Tree and Neural Network precision dropped nearly 29\% and 30\% respectively. This attack was triumphant in degrading the performance of the HMD to almost 50\% precision, thereby eroding the trust in the system. }

\section{Related Work}
\label{sec:rel}

\subsection{Hardware Performance Counter-based Malware Detection}
Malware tend to compromise the security of a system by modifying, damaging, or gaining unauthorised access without the knowledge of the user. Effects of these Malware can range from stealing sensitive information, modifying system's functionality as well as performing Denial of Service (DoS) attacks. In order to overcome the challenges of AVS, trusted Hardware-based Malware Detection (HMD) techniques are being proposed. These techniques tend to use low level micro-architectural features in a processor, like HPCs, to differentiate between malicious software and their benign counterparts. 

One of the preliminary methods to detect Malware, using HMDs based on HPCs was proposed in \cite{malone2011hardware}. \cite{demme2013feasibility} developed an algorithm that uses data from the existing HPCs and applies ML techniques, namely Artificial Neural Network and K-Nearest Neighbour, to classify between Malware and benign applications. A virtual machine monitor, NumChecker, has been presented in \cite{wang2015reusing} to measure system call events from HPC values in order to detect malicious kernel control-flow modifications. Another approach of detecting anomaly in a system has been introduced by \cite{krishnamurthy2019anomaly}, which does real-time quantification of HPCs during execution to detect Malware in  cyber-physical systems. In \cite{wang2016malicious}, a low-cost HPC-based Malware detection technique, ConFirm, has been proposed, which uses comparison-based approach to detect firmware modifications with simple control flows. \cite{jyothi2016brain} approached this domain by designing Behaviour-based Adaptive Intrusion detection in Networks (BRAIN). The proposed approach uses HPCs to combine network statistics and modeled application behaviour to detect Denial of Service (DoS) and Distributed Denial of Service (DDoS) attacks. Similar HPC-based Malware detection techniques have been discussed in \cite{bahador2014hpcmalhunter,wang2016hardware,tang2014unsupervised,ozsoy2015Malware}, which protect the vulnerable systems from hostile software programs and leaves the attacker with an onerous job of tampering the trusted hardware architecture. Beyond HPCs, other HMDs were developed by \cite{zhou2018hardware,zhou2016hardware,zhou2017hardware,zhou2016hardware1}.

A lightweight node-level Malware detector and network-level Malware confinement algorithm have been proposed in \cite{dinakarrao2019lightweight}, which uses a runtime Malware detector \textit{HaRM} to detect vulnerabilities based on HPC values on an IoT network architecture. \cite{sayadi20192smart} developed an alternative HMD approach by introducing \textit{2SMaRT}, a two stage ML-based classification algorithm, which selects the best HPCs by feature selection, and feeds them into the bi-stage classifier to efficiently detect Malware applications. \textcolor{black}{In \cite{khasawneh2017rhmd}, the authors had proven that HMDs trained on dynamic traces can be successfully reverse engineered and evaded through code injection in a malicious program. Furthermore, the authors proposed Resilient HMDs (RHMDs) which randomly switch between different detectors to improve the resiliency to reverse engineering attempts. However, as claimed by \cite{dinakarrao2019adversarial} and shown in  \cite{goodfellow2014explaining,liu2016delving, papernot2016limitations, szegedy2013intriguing}, carefully crafted perturbations into the input data can modify the classifier output. Adversarial attacks on such Malware classifiers have been proposed in \cite{dinakarrao2019adversarial}, which, unlike \cite{khasawneh2017rhmd}, uses adversarial learning to alter the HPC values of a program in parallel to its' execution on the same thread in order to deceive the ML classifiers of the HMD, thereby overlooking a Malware that can potentially disrupt a system on the go. In this paper, we propose a MTD-based approach that defends against such adversarial attacks.}

\subsection{Moving Target Defense (MTD)}
With the increasing threat to sensitive information, identifying and removing security vulnerabilities are emerging as need of the hour for modern computing systems. The never-ending game between the attackers and defenders is what pushes the development of new defense techniques to address the cyber-threats and thwart the attackers' ploy. MTD is one such promising cyber-defense strategy, that can potentially challenge the attack capability on cyber physical systems. This dynamic defense technique, originally conceived by the Department of Homeland Security \cite{3reasons17:online}, constantly drives the attack surface to evolve across multiple system dimensions. Reducing or changing the attack surface dynamically confuses the planned attack, decreasing the system's susceptibility. 

A fundamental MTD architecture called Mutable Networks (MUTE) has been proposed in \cite{al2011toward}, enabling systems to dynamically change network configurations by host IP mutation, in order to jeopardize adversarial attacks. Similar network randomisation approaches have been proposed in \cite{antonatos2007defending, dunlop2011mt6d}, which slows down the hitlist worms on attacking the system. \cite{shacham2004effectiveness, kil2006address} approaches the dynamic defense strategy by randomly shifting critical memory positions of certain system components, thereby converting a malicious cyber attack into a benign process crash. Instruction set randomisation-based MTD approach has been proposed in \cite{boyd2008general} for safegaurding networks against code injection attack. To create diversified attack surfaces for internet services, a software diversification-based dynamic defense model has been proposed in \cite{huang2011introducing}, which creates a set of random virtual servers by configuring it with a unique software mix, thereby complicating the attack.

\section{Moving Target Defense}
\label{sec:MTD}
\subsection{Motivation}
\label{sec:MTD:Motivation}
\begin{figure}[h!]
  \centerline{\includegraphics[scale = 0.4]{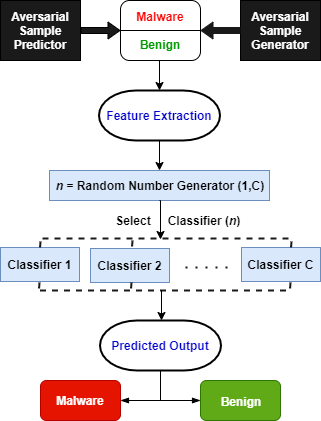}}
  \caption{Proposed Moving Target Defense.}
  \label{fig:mtdpicture}
\end{figure}

The adversarial attack in Section~\ref{sec:background-adversarialattack} raises questions on the credibility of the HMD in identifying a potential Malware from the harvested HPC traces. Any modern processor supports multiple HPCs (up-to few hundreds \cite{singh2017detection}). During an adversarial attack, only specific HPCs are being targeted by an attacker. Therefore, a defense that constantly changes the attack surface by modifying the HPC combinations possible is robust in countering this adversarial attack. The proposed algorithm perpetually modifies the following parameters employed in Malware detection \textemdash $(1)$ the count of HPCs harvested from the processor, $(2)$ the specific set of HPCs to be provided as input to the HMD, and $(3)$ precisely, which classifier of the HMD is utilising which HPC combinations. A process flow diagram for proposed MTD is shown in Figure~\ref{fig:mtdpicture}. 

Since, the MTD algorithm is contingent on multiple ML classifiers with distinct HPC combinations, it is crucial to determine which HPC is placed in which group, and eventually, in which classifier. Section \ref{sec:featuretesting} deals with this outline of feature testing which returns the best performing HPCs among those in a processor. Feature selection explains the concept of clustering the HPCs based on the respective results from the previous step of feature testing, as explained in Section \ref{sec:featuretesting-featureselection}. Section~\ref{sec:featuretesting-analysismtd} presents the proposed MTD algorithm with multivariate HMD classifiers. Finally, in Section \ref{sec:MTD:CCA}, we develop an analytical model to analyse the robustness of our MTD algorithm against the adversarial attack framework trying to reverse engineer the HMD.

\subsection{Feature Testing}
\label{sec:featuretesting}
With a plethora of possible combinations for generating potential ML classifiers with unique HPC clusters, the best HPCs need to be utilized in order to obtain the highest accuracy. Feature testing is used to address this challenge.
There are three different feature testing techniques employed to assist in producing ML models with good accuracy, which are briefly explained, as follows:

\subsubsection{Univariate Selection}
\label{sec:featuretesting-uni}
Univariate selection is the usage of statistical tests that will assist in computing the features having the best correlation with the output. For the experiments conducted, {\it python scikit-learn} library was used, allowing access to select K\footnote{K is an user-defined positive integer.} best classes that were executed with the chi-squared statistical test to ensure positive HPC values. Assuming $H$ to be the set of all HPCs in a system, each individual HPC in $H$ will be harvested from executing multiple applications. The data is used as input to the univariate selection testing, furnishing a set of HPCs with high interdependence on output.

\subsubsection{Feature Importance}
\label{sec:featuretesting-featureimp}
Feature importance provides a score for each of the input features (HPCs in our case), that represents how significant a specific feature is towards the output variable. A higher score establishes which features will be more relevant and therefore, more prominent in incorporating in a ML classifier. Feature importance uses a decision tree classifier model that eradicates unnecessary branches as it is trained. This ensues a reduction in redundant data and decision making from noise resulting in improved accuracy. Similar to univariate selection, the feature importance of each HPC is obtained by executing a number of programs. The data is then used as input to the feature importance testing which will return the scores for each HPC. A higher score signifies a HPCs being more dominant in Malware detection. 

\subsubsection{Correlation Matrix with Heat map}
\label{sec:featuretesting-correlationwithheatmap}
In ML models, a correlation matrix explains how the features relate with others or a specified variable. Correlation matrices mapped onto a heat map are effective in identifying the features with the least and most impact on a specific variable. The heat map utilized is color coded such that blocks that are more green have higher correlation while blocks that are more red have lower correlation. Correlation can be used to understand whether a feature increments or decrements the value of the output variable. The correlation matrix obtained from the HPC data for several applications will provide a heat map showing the impact of each feature in the system. 

\begin{figure}[h!]
\centering
  \includegraphics[width=0.9\linewidth]{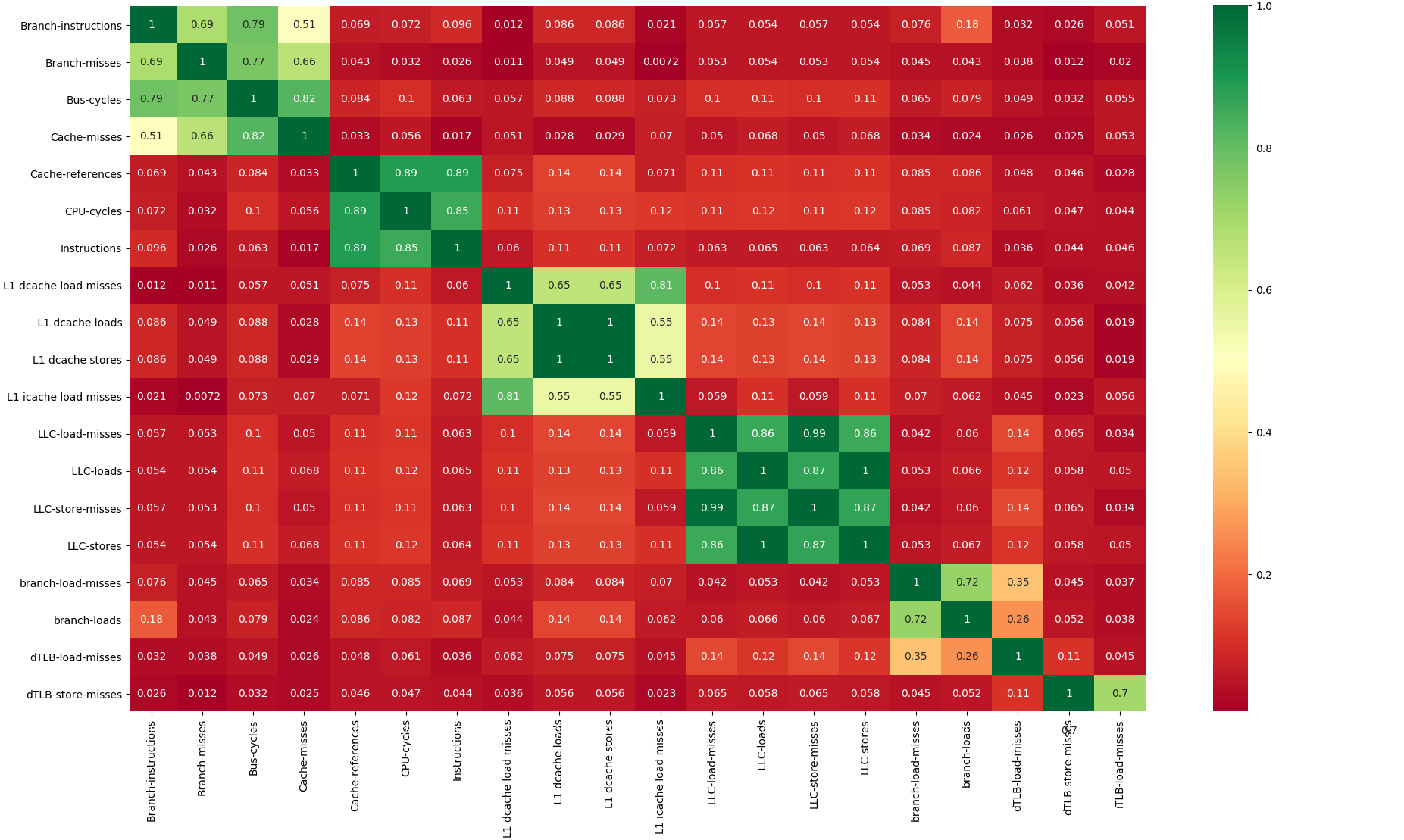}
  \caption{Heat map for HPCs in Raspberry Pi 3.}
  \label{fig:heatmap}
\end{figure}

\subsection{Feature Selection}
\label{sec:featuretesting-featureselection}
The three feature testing techniques described in Section~\ref{sec:featuretesting} can now be applied in building the MTD. An analysis can be preformed to determine the best HPC combinations for training a classifier from the results of the three techniques. 
Since the goal of our proposed MTD is to create multiple ML classifiers while maintaining acceptable accuracy, the best HPCs should be paired with HPCs that are average to ample. The adversarial attack that is being defended against specifically targets branch-misses and LLC-load-misses. Therefore, HPCs that are highly correlated can be paired up with these HPCs that are being targeted to try and bolster the accuracy of classification. 

The heat map for the HPCs utilized in the experiment on {\it Raspberry PI 3} is shown in Figure~\ref{fig:heatmap}. The more green a square block is, the more correlation that HPC has towards the corresponding HPC. Additionally, each square block has a value between 0 and 1. The higher the value is, the more correlation it has and vice versa. Based on the heat map, potential HPC combination pairs can be produced. For example, from the heat map, HPCs cache-references, CPU cycles, and instructions have high correlation with each other. This is proven by the fact that they're are all green when corresponding with each other while they are red when corresponding with other HPCs. Comparing univariate selection and feature importance results validate that these HPCs have high correlation with the output. Therefore, it would be ideal to combine these HPCs together in a ML classifier to be utilized by the MTD. The employment of these three feature testing techniques produces high performing ML classifiers for MTD.

\subsection{Analysis of Moving Target Defense}
\label{sec:featuretesting-analysismtd}
Once the best and optimal number of classifiers and HPCs are selected, the MTD can be modelled. The MTD evaluates the HPCs of a specific application (benign or malicious software) and classifies them accordingly. Because HPC data are collected in specific intervals (e.g. 10 ms), the HPC data will be spilt into individual iterations. For each iteration, a random number will be generated between a range of (1, N) inclusive where N is the number of classifiers being utilized in the MTD. The value of the random number generated will determine which classifier will receive that HPC iteration data. This specific classifier will predict whether the given HPC iteration data corresponds to benign or malicious. Hence, every time the HPC data is sent to the MTD, the classifiers utilized will constantly be modified. Continually changing the attack surface is the fundamental concept of the MTD. The MTD inherits its name due to the fact that the HMD classifier used to predict whether the program is benign or malicious is perpetually altering, thereby obscuring the attacker from the set of HPCs that the MTD will be evaluating on. This entire MTD-based Malware detection approach is described in Algorithm 2.
For experimental purposes, to test and verify the MTD implementation, every predict value returned was checked to report the accuracy.

\begin{algorithm}[h!]
\label{sec:algorithm5}
\vspace{1mm}
\textbf{Input}: Applications (Malware/Benign), k number of features to be returned from univariate selection, C number of classifiers to be created\\
\textbf{Output:} Number of ML Classifiers, MTD Accuracy  \\
\vspace{-4mm}
\caption{Moving Target Defense Design and Application}
\begin{algorithmic}[1]
\State \textbf{design\_of\_classifiers()}\{
\State \textbf{\#define} int $count\_of\_classifiers$ = 0
\State {\it data} = Get HPC Values from Applications
\State {\it univar} = UnivariateSelection({\it data},{\it k})
\State {\it featureimp} = FeatureImportance({\it data})
\State {\it hmap} = heatmap({\it data})
\While {{\it count\_of\_classifiers} $<$  {\it C}}
\State {\it h'} = obtain high correlation HPC Set from {\it hmap}
\State \textbf{for} high $(correlation\_values)_{\it h'}$ in {\it hmap}
\State \hspace{4mm} \textbf{if} $(univar)_{\it h'}$ and $(featureimp)_{\it h'}$ is high
\State \hspace{8mm} {\it c(h')} = create classifier with {\it h'}
\State \hspace{8mm} {\it C} = {\it C} + {\it c(h')}
\State \hspace{8mm} {\it count\_of\_classifiers } = {\it count\_of\_classifiers} + 1
\State {\it hmap } = {\it hmap} - {\it h'}
\EndWhile
\State \textbf{output} ML\_classifiers({\it C})
\State \textbf{end}
\State \}
\State \textbf{application()}\{
\State \textbf{\#define} int $pass,fail$
\State \text{\textit{input}} = read testing data
\While{input}
\State \textit{n} = RandomNumberGenerator (1, C)
\State \textbf{if}  $classifier_n$ \hspace{5mm}
\State \hspace{8mm}\{$predict_n$ ({\it input})\}
\State \textbf{if} \textit{$predict_n$} passed :
\State \hspace{8mm}\{{\it pass}++\}
\State \textbf{else}\hspace{1.5mm} \{{\it fail} ++\}
\EndWhile
\State \textbf{output} MTDaccuracy = ({\it pass} / ({\it pass} + {\it fail}))
\State \textbf{end }
\State \}
\end{algorithmic}
\end{algorithm}

\subsection{Mathematical analysis of the proposed MTD}
\label{sec:MTD:CCA}
The proposed MTD model uses multiple ML classifiers trained on different HPC clusters. As discussed in Section~\ref{sec:background-HPCinML}, the number of HPCs varies depending on the processor. In addition, based on the processor, only a few HPCs can be monitored simultaneously. The minimum number of HPCs corresponding to each classifier used in HMDs is 1, maximum being the number of HPCs that the CPU allows to observe simultaneously. 

In this section, we will explain the possible classifier combinations that the model can obtain, given the number of total HPCs available as well as the maximum number of HPCs that a ML classifier can accommodate, using an analytical framework. For simplicity, we assume only one type of ML model, i.e., either Decision Tree or Neural Network. The computation becomes even more complex when different types of models are considered together. Let $H_{t}$ be the total number of HPCs in the processor, and $X$ be the number of counters that is used for a ML classifier. As stated earlier, the model has constraints on the number of HPCs that can be fitted per classifier, with the minimum value being 1. 
 Let $R_{max}$ be the maximum number of HPCs that can be monitored simultaneously for a particular processor. Thus, the value $X$ ranges from a minimum of 1 to a maximum of $R_{max}$, all inclusive. The cumulative number of different types of classifiers, each having unique set of HPC values can be represented by the following Equation:  

\begin{equation}
\label{equ5}  
Total \ Classifiers  = {H_{t} \choose 1} + ... + {H_{t} \choose R_{max}} 
\end{equation}

Equation~\ref{equ5} can be generalized as:

\begin{equation}
\label{equ6}
 Total \hspace{1mm} Classifiers\ (N_{h}) = \sum_{i=1}^{R_{max}}\hspace{1mm} {H_{t} \choose i}
\end{equation}

where the total number of classifiers with unique HPC combinations is denoted by $N_{h}$. $X$, $H_{t}$ and $R_{max}$ are all positive integers, where $R_{max}$ typically varies between $[4,7]$ for all modern processors.

For a \textit{Raspberry Pi 3} with ARM Cortex-A53 CPU, the value of total number of available HPCs, $H_{t}$ is $20$, and the value of $R{max}$ is considered to be 4. Therefore, the possible cumulative classifier models with distinct set of counters for such a processor is calculated from Equation~\ref{equ6} as:

\begin{equation} 
\label{equ7} 
N_{h} =  {20 \choose 1} + {20 \choose 2} + {20 \choose 3} + {20 \choose 4} = 6,195
\end{equation}

\begin{figure}
\centering
  \includegraphics[width=0.9\linewidth]{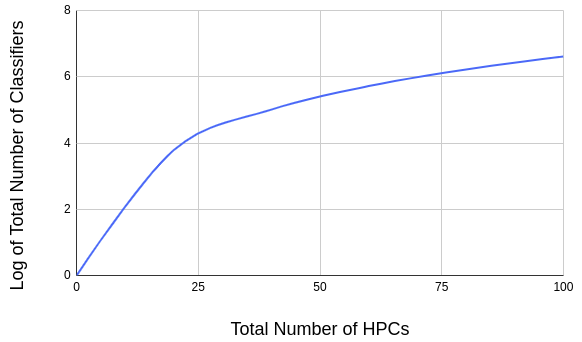}
  \caption{Variation of the number of cumulative classifiers with total HPCs.}
  \label{fig:hpcvsclass1}
\end{figure}

Thus, there are 6195 potential classifiers to be used for the proposed MTD model consisting of various HPC combinations. With the linear increment in the number of HPCs, $H_{t}$, the total number of classifiers with distinct counter combinations $N_{h}$ increases drastically. Figure~\ref{fig:hpcvsclass1} represents the non-linear variation in the number of the ML classifiers with the total number of accessible HPCs in a processor. For a CPU with hundreds of HPCs, this value exceeds more than  $4\times10^{6}$.

Now, we will analyse the different ML classifier combinations possible for the MTD model, given a particular value of $H_t$ and $R{max}$. From Equation~\ref{equ6}, we can obtain the number of classifiers for such a combination as $N_{h}$. A defender can choose any number of classifiers from this set.  Let $C$ be a number of classifiers that can be selected from the set of $N_{h}$ classifiers. The MTD model behaviour requires at least 2 classifiers. The maximum number of classifiers that the MTD model can choose is $N_{h}$. Therefore, the value of $C$ is constrained to $2 \leq C \leq N_{h}$. The aggregate of the number of such classifiers that can be picked together for MTD model can be represented by :

\begin{equation} 
\label{equ8}
Total\ Combinations \ (N_{c})= {\ N_{h} \choose 2} + \ldots + {\ N_{h} \choose \ N_{h}} \end{equation}

where $N_{c}$ denotes the total number of classifier combinations, and $N_{h}$ is the total number of classifiers with distinct HPC clusters, $N_{c}$ and $N_{h}$ being positive integers.

Equation~\ref{equ8} can be rewritten as :

\begin{multline}
\label{equ9}
  N_{c} = \Bigg [ {N_{h} \choose 0} + {N_{h} \choose 1} +{N_{h} \choose 2} + {N_{h} \choose 3} + \ldots + \\  {N_{h} \choose N_{h}-1} + {N_{h} \choose N_{h}} \Bigg ] - {N_{h} \choose 0}  - {N_{h} \choose 1}
\end{multline}

Simplifying this using a Binomial expression, we obtain:

\begin{equation}
\label{equ10}
  N_{c} = (1+1)^{N_{h}} - {N_{h} \choose 1} - {N_{h} \choose 0} \\
  = 2^{N_{h}} - N_{h} - 1
\end{equation}

Equation~\ref{equ10} furnishes the total number of ML classifier combinations $N_{c}$ for the MTD, given the total number of classifiers $N_{h}$, obtained from Equation~\ref{equ6}.

\begin{figure}
\centering
  \includegraphics[width=0.9\linewidth]{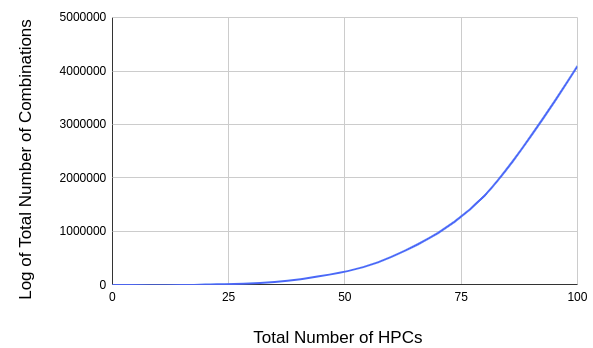}
  \caption{ Number of Possible Combinations based on Variation of Total HPCs.}
  \label{fig:hpcvscombo}
\end{figure}

As seen from Equation~\ref{equ7}, for a \textit{Raspberry Pi 3} with $20$ HPCs, the total number of classifiers $N_{h}$ is $6,195$. Using the $N_{h}$ value in Equation~\ref{equ10}, we obtain  approximately $7.6\times10^{1864}$ combinations that the defender can use to model a MTD. Figure~\ref{fig:hpcvscombo} shows the variation of the number of classifier combinations against the total number of HPCs present in a CPU, where $N_c$ increases exponentially with increase in $H_t$.

These extensive classifier combination values eventually establish the defense against the attacker trying to reverse engineer the classifiers. An adversary attempting to break into the system has to know exactly what HPCs a particular classifier is using. But, with the dynamic attack surface variation established by the MTD, it is nearly impossible for the attacker to extract the relevant attack information; the probability of which is $\frac{1}{7.6\times10^{1864}}$,  which is approximately $0.13157\times10^{-1864}$, for even a small processor with 20 HPCs. This probability is not contingent on what classifier algorithm the MTD model is utilising, thereby yielding a consistent defense strategy against the adversarial attacks by complicating the detection model.
\textcolor{black}{In contrast to our proposed MTD, if we consider to have a single classifier with a specific number of HPCs, the probability of such a system to succumb to reverse engineering attack by an adversary is significantly increased. Let us consider a situation, where the Malware detector employs a single classifier with a fixed number of HPCs ($h$) for classifying an application. Now, with $h$ HPCs among $H_{t}$ ($H_{t}$ being the total number of HPCs available in the system), an attacker can choose between $H_{t} \choose h$ combinations of HPCs. Among these, only one combination is correct; hence the probability to successfully guess the employed set of HPCs is 1/$H_{t} \choose h$. For example, in a processor with a total of 20 HPCs, if the HMD employs eight HPCs for Malware detection, the adversary can choose between $20 \choose 8$ $= 125970$ combinations, out of which, the probability of choosing the correct combination is $1/125970$ = $0.793839 \times 10^{-5}$. This is significantly large, when compared to the probability $0.13157\times10^{-1864}$, as obtained from the dynamic variation of attack surface in the MTD. Hence, incorporating multiple classifiers, as opposed to one bolsters the Malware detector remarkably, against the aforementioned adversarial attack. }

\section{Experimental Results}
\label{sec:Experiment}

\subsection{Experimental Setup}
\label{exp:setup}
Recently, \cite{das2019sok} examined some perils of using HPCs for Malware detection, in which, the authors argued against the use of Virtual Machines (VM) for designing the HMD, since HPCs in a VM differ from those in a bare-metal. Therefore, for developing a proof of concept of our proposed MTD methodology, we eradicated the need for a VM by conducting tests on a {\it Raspberry Pi 3 Model B}, which utilizes an ARM processor, and HPC values were collected using {\it perf 4.18}. Based on the data in Figure~\ref{fig:classpre} and Figure~\ref{fig:classacc}, the ML models utilized in the MTD were Decision Tree (DT) and Neural Network (NN) classifiers. These ML models were written in {\it Python3} using {\it scikit-learn}. The Malware samples were acquired from Virusshare \cite{VirusSha57:online}, and the benign programs include the MiBench \cite{MiBenchv5:online} benchmark as well as different sorting and computational algorithms. 300 Malware and 300 benign programs were executed on the {\it Raspberry Pi}, and the HPC values were harvested. For each HMD, a classifier was trained on HPC branch-instructions, HPC branch-misses, HPC instructions, and HPC LLC-load-misses, which are the HPCs used by \cite{dinakarrao2019adversarial}. Training these classifiers on the HPCs allows us to obtain the initial accuracy before the adversarial attack.

The adversarial attack for branch misses and LLC-load-misses were written as C code, which were then ran together with the programs to be camouflaged through a python wrapper. 50 new Malware samples, which were not included in the training data set, were tested on each trained classifier to prove that the accuracy would go down. These 50 Malware were first tested on each classifier with no HPC perturbations to show that the ML models trained could label them with satisfactory accuracy. Subsequently, these 50 Malware samples were executed again with the adversarial HPC generator, and each classifier was tested to demonstrate reduction in accuracy.

Finally, the 50 programs with HPC perturbations were executed through the proposed MTD to prove that the defense could recover the lost accuracy. 
In the MTD, two classifiers using the same ML algorithm were implemented using HPCs Branch-instructions, Branch-misses, Bus-cycles, and Cache-misses for one classifier, and HPCs Cache-references, CPU-cycles, and Instructions for the other classifier. These HPCs were picked based on the heat map in Figure~\ref{fig:heatmap}, followed by further analysis using other feature selection methods described in Section IV-C, which shows that these HPCs have good correlation with each other. Two different versions of the MTD were tested where the only difference was the ML algorithm used for the two classifiers. 

\subsection{Results}
\label{exp:results}

\begin{figure}[h!]
\centering
  \includegraphics[width=0.9\linewidth]{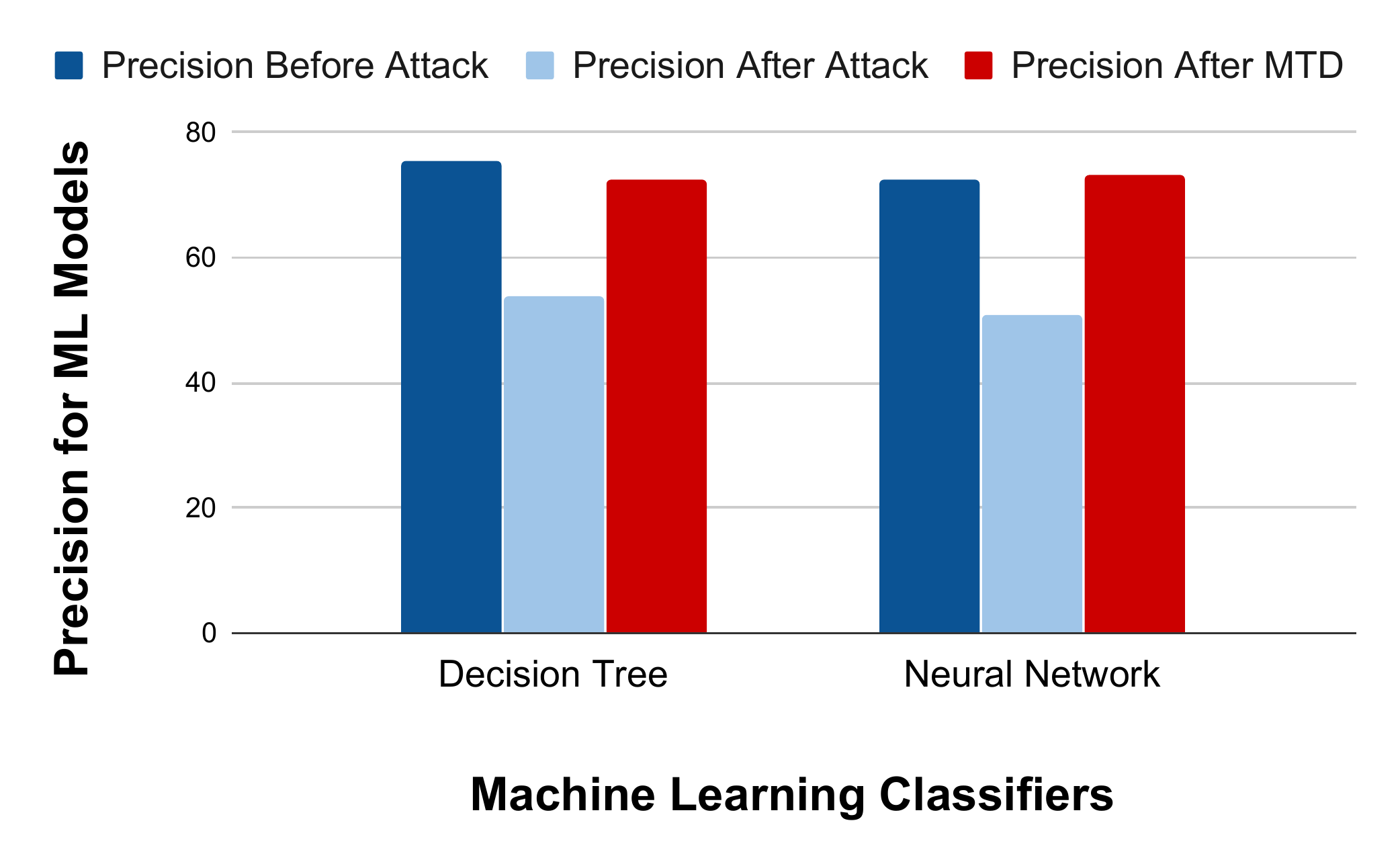}
  \caption{Precision with Moving Target Defense.}
  \label{fig:precaftermtd}
\end{figure}

\begin{figure}[h!]
\centering
  \includegraphics[width=0.9\linewidth]{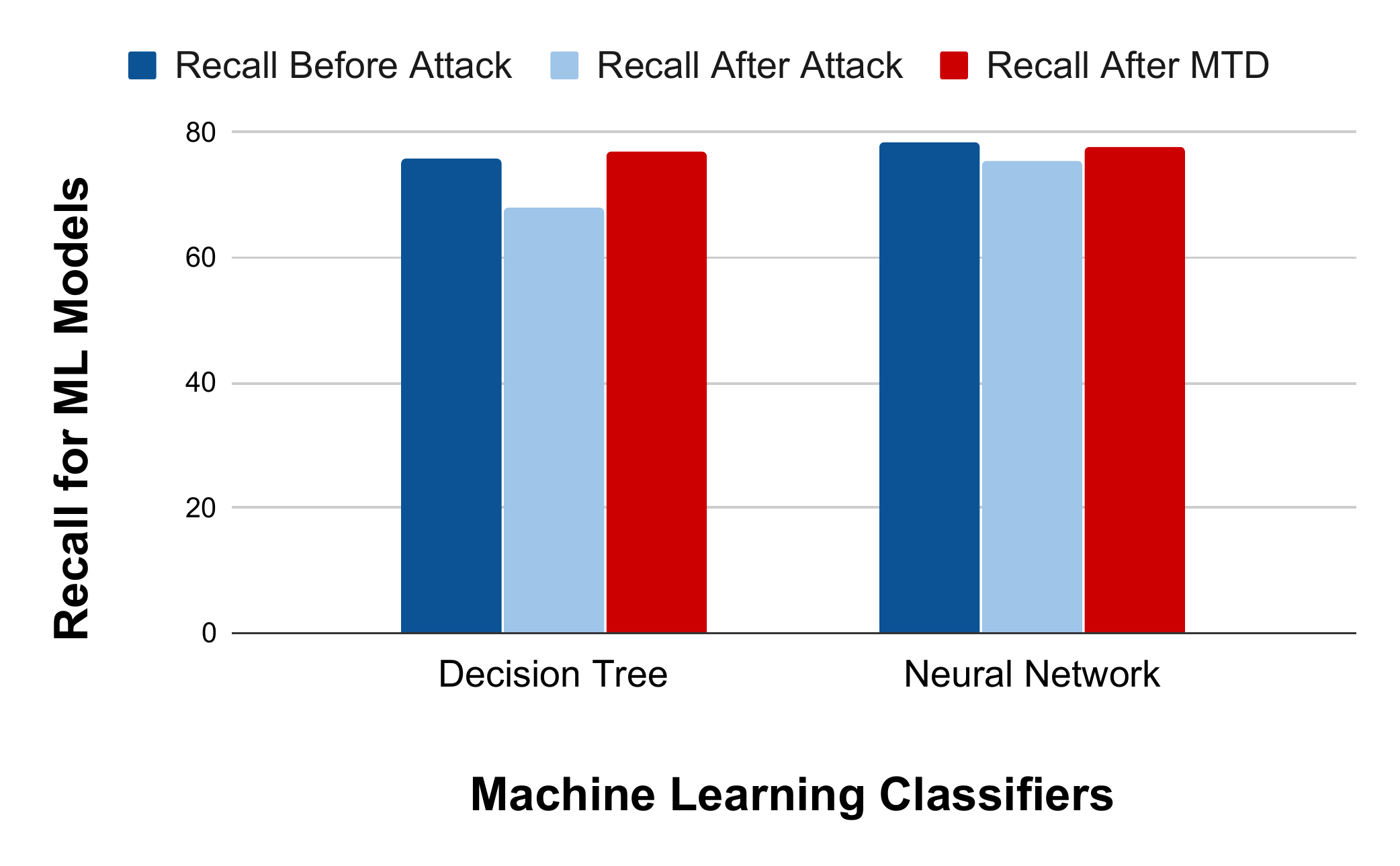}
  \caption{Recall with Moving Target Defense.}
  \label{fig:recaftermtd}
\end{figure}

\begin{figure}[h!]
\centering
  \includegraphics[width=0.9\linewidth]{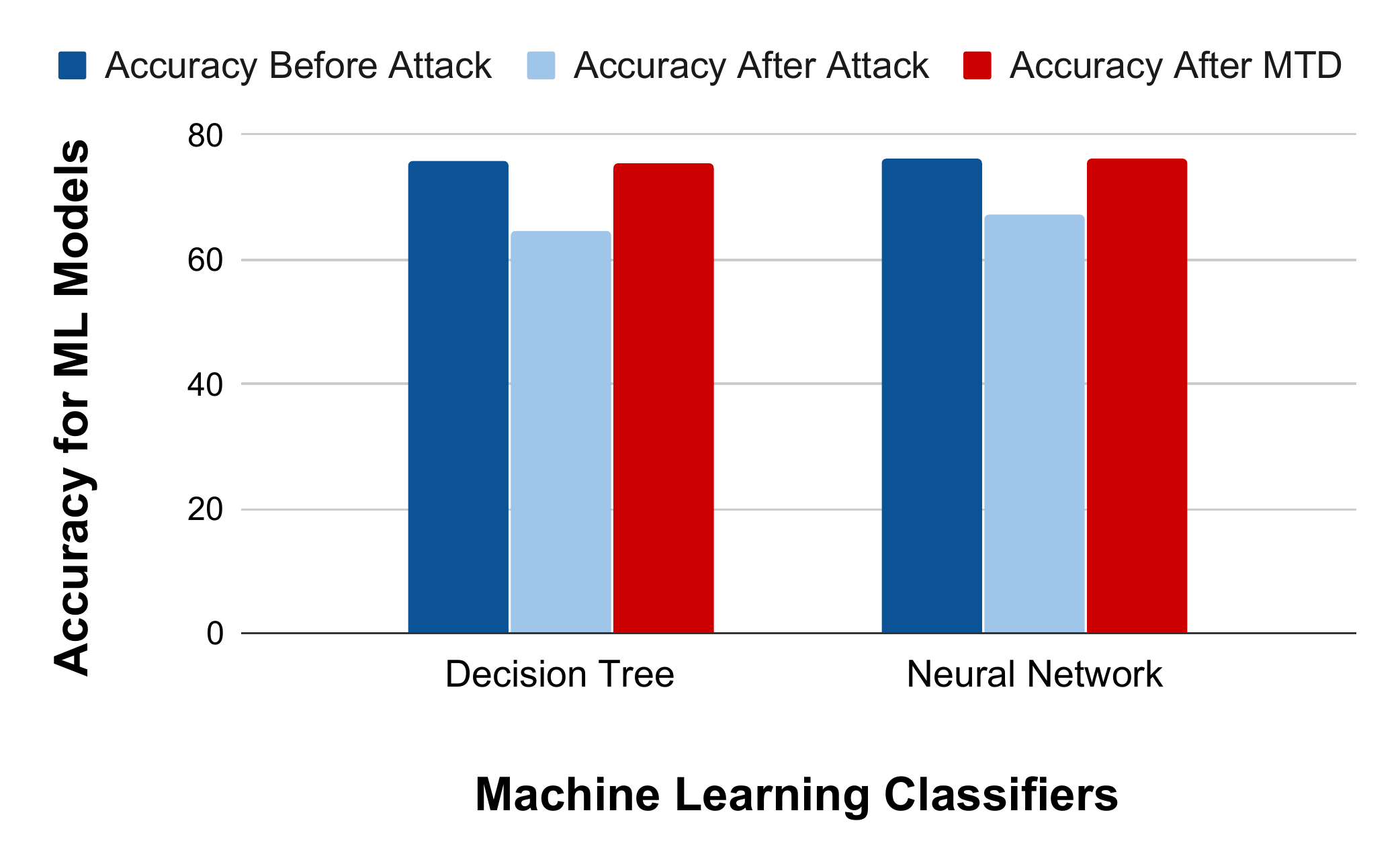}
  \caption{Accuracy with Moving Target Defense.}
  \label{fig:accaftermtd}
\end{figure}

\textcolor{black}{Our predominant concern is the MTD's capability to identify the malicious samples. Therefore, we are primarily interested in the precision of the MTD. The initial training precision, without any perturbations, for the DT and NN classifiers was 75.65\% and 72.43\%, respectively. The Malware detection precision for the 50 Malware programs with no perturbations was 74.2\% and 72.1\% for DT and NN, respectively. This shows that each classifier has competent precision in properly classifying the applications. The precision for the 50 Malware with perturbations in HPC traces for each classifier was 53.96\% for DT and 50.96\% for NN which is expected as the attack should reduce the precision. However, the precision for the 50 Malware with perturbations running through the MTD was 72.42\% using DT and 73.22\% using NN. To obtain other performance metrics, the 50 Malware with perturbations was tested along with 50 new benign programs. Comparison of precision, recall, and accuracy for the two ML classifiers for each of the three cases, i.e., before attack, after attack, and after MTD are shown in  Figure~\ref{fig:precaftermtd}, 
Figure~\ref{fig:recaftermtd}, and Figure~\ref{fig:accaftermtd} respectively. The 50 benign programs bolsters the accuracy and recall from plummeting as much as the recall because the adversarial attack only functions on Malware. It is observed that the adversarial attack reduced the precision drastically to almost 50\%, which is as best as random guess. However, the proposed MTD technique was successful in improving the precision, by nearly 18.46\% for DT and 22.6\% for NN. Some of the performance metrics after the MTD are greater than the base classifier because some ML models utilized in the MTD are superior, engendered from our feature selection usage described in Section~\ref{sec:featuretesting-featureselection}.  }

\subsubsection{Efficiency of MTD on Real-life Applications}
\label{datasetbolsteredinpaper}
\begin{figure}[h!]
\centering
  \includegraphics[width=0.9\linewidth]{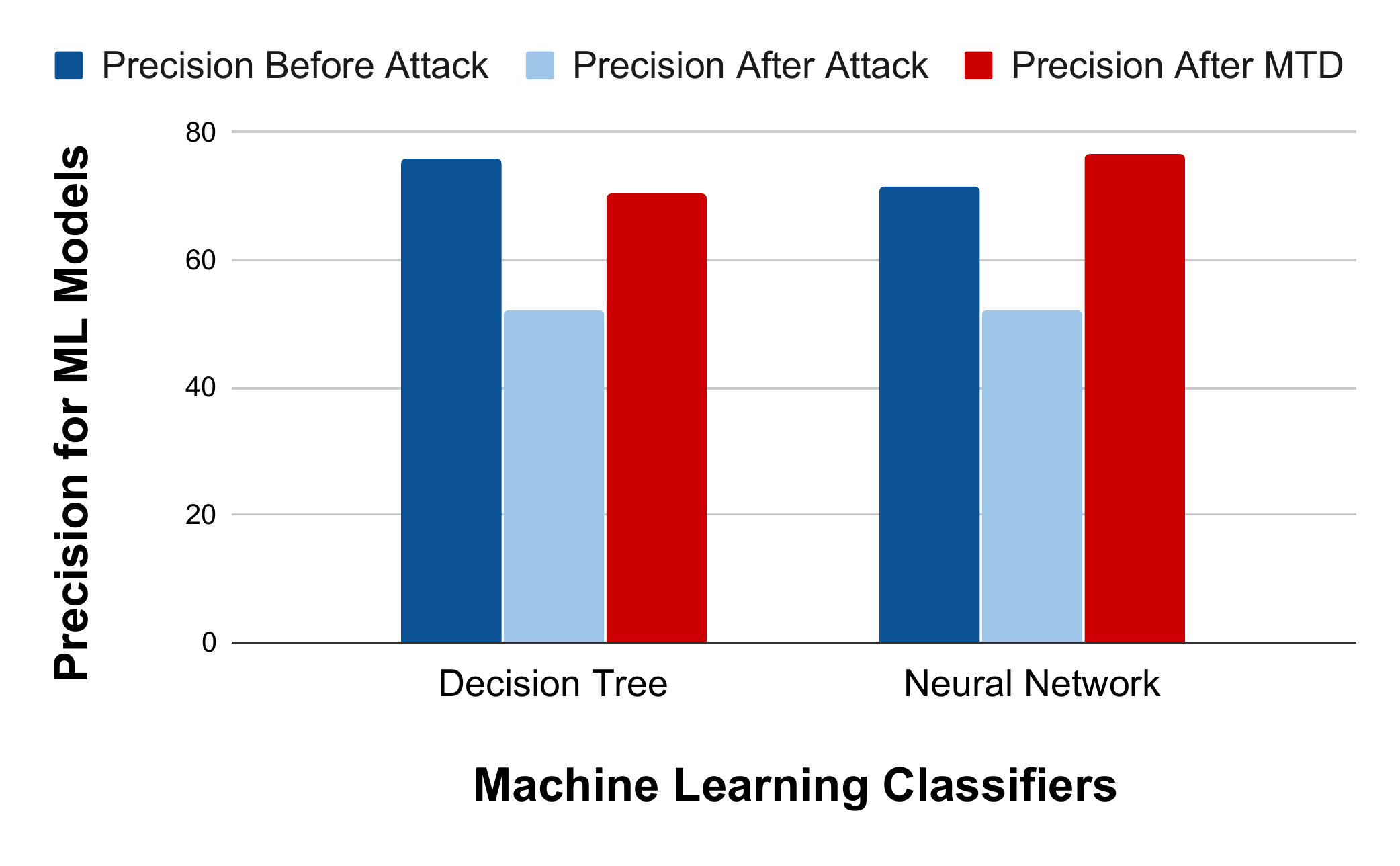}
  \caption{Moving Target Defense on Real-life Applications.}
  \label{fig:precaftermtdnewdatasetinpaper}
\end{figure}

\textcolor{black}{
While the MTD was able to recover the precision after the adversarial attack in Section~\ref{exp:results}, the lack of real-life applications in our dataset induces skepticism of the security and practicality of the defense. In this experiment, we have adjusted our training set by incorporating real-life applications such as web browsers, word processors, and programming IDEs. Furthermore, we have also bolstered our data set by utilizing the SPEC and Phoronix benchmarks, which represent regular processor workloads \cite{SPECBenc57:online,OpenBenc36:online}. To demonstrate the robustness of the MTD, the experiment in Section~\ref{exp:results} was reproduced. We utilized the same HPCs for our baseline models and our MTD models.  }
\textcolor{black}{
The initial precision for the models was 75.79\% and 71.67\% for DT and NN. The Malware detection precision for the 50 Malware programs with no perturbations was 75.78\% and 71.2\% respectively. Subsequently, these 50 Malware were executed again with the adversarial HPC generator and tested along with 50 additional benign programs, on the classifiers again. The precision for the DT dropped to 52\%, and the NN reduced to 51.98\%. However, these same Malware with perturbations and benign programs were ran through the MTD furnishing a precision of 70.52\% and 76.82\%. Our experimental results for our adjusted dataset are shown in Figure~\ref{fig:precaftermtdnewdatasetinpaper}.
}

\subsubsection{Impact of number of classifiers used in MTD}
\label{classacctradeoff}
\begin{figure}[h!]
\centering
  \includegraphics[width=0.9\linewidth]{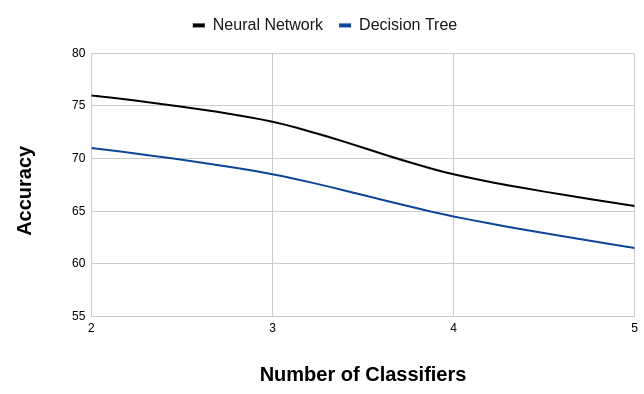}
  \caption{Classifier versus Accuracy Trade-off.}
  \label{fig:classvsacc}
\end{figure}

\textcolor{black}{We previously incorporated benign programs in our data set in order to attain a recall metric. In the impending experiments, we only utilized Malware programs because we are primarily focused on the Malware detection capabilities of the MTD, i.e., the precision is the principal performance metric. Since there are no benign programs involved in these experiments, the accuracy and the precision are equivalent.} Figure~\ref{fig:classvsacc} depicts the variation in classification accuracy when the number of classifiers change. The MTD was tested using the 50 Malware with perturbations, but the number of classifiers was incremented in each round. The additional classifiers were chosen using methods described in Section~\ref{sec:featuretesting} and Section~\ref{sec:featuretesting-featureselection}. As the number of classifiers increased, the accuracy diminished. This is because, certain HPCs are better correlated than others for training the ML models. Therefore, assuming HPCs cannot be repeated; as the number of classifiers increases, the more non-optimal HPCs are integrated into the MTD. Consequently, the accuracy will decrease. However, the defense will improve considering utilization of additional classifiers results in attackers having an arduous time in reverse engineering the system as they will not know which HPCs the ML classifiers are using, as evident from Equation~\ref{equ10}. From Figure~\ref{fig:classvsacc}, the MTD NN was reduced from 75.98\% accuracy to 65.5\% as the number of classifiers was increased to five. The MTD DT accuracy was reduced from 71\% to 61.5\% as the number of classifiers was incremented to five. Thus, selecting just two classifiers for MTD furnishes the best accuracy.

\subsubsection{Classifier Priority}
\label{classprio}
\begin{figure}[h!]
\centering
  \includegraphics[width=0.9\linewidth]{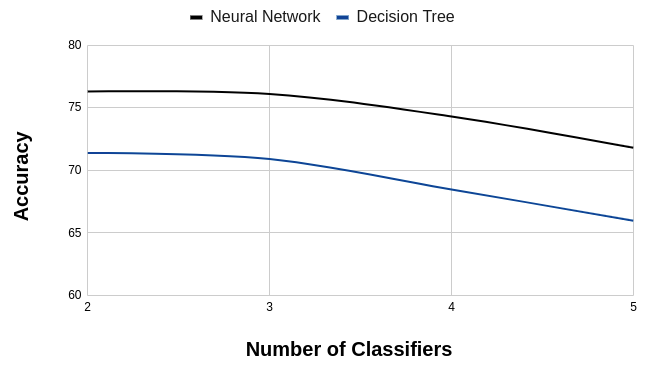}
  \caption{Assigning Priority to the Best classifier.}
  \label{fig:classvsaccprio2x}
\end{figure}

 In this experiment, similar to Section~\ref{classacctradeoff}, five classifiers were chosen. The MTD was tested using the 50 Malware with perturbations, but instead of random classifier selection, the best classifier was given priority in the MTD. Similar to Section~\ref{classacctradeoff}, variation of accuracy with an increase in the number of classifiers is shown in Figure~\ref{fig:classvsaccprio2x}. In this case, the best classifier was utilized every other run, while the other classifiers are chosen randomly using MTD, from the set of the remaining four classifiers. Figure~\ref{fig:classvsaccprio2x} shows that while the accuracy still reduces as the number of classifiers is incremented, the accuracy doesn't plummet as quickly as in Figure~\ref{fig:classvsacc}. The classifier with priority is able to increase the accuracy since there is less testing data being sent to classifiers with inferior accuracy. The MTD DT accuracy reduced from 71\% to 65.8\% as the number of classifiers was incremented to five. The MTD NN accuracy reduced from 75.98\% accuracy to 71.5\% as the number of classifiers was increased to five. Compared to Figure~\ref{fig:classvsacc}, where Decision Tree furnished 61.5\% and Neural Network furnished 65.5\% accuracy with five classifiers, it is demonstrated that selecting the classifier with highest priority can improve the classification accuracy. Hence, if more than two classifiers are used for the MTD, it is better to use a priority-based selection.

\subsubsection{Mixing Classifiers}
\label{mixingclass}
\begin{figure}[h!]
\centering
  \includegraphics[width=6.5cm,height=5cm,angle=0]{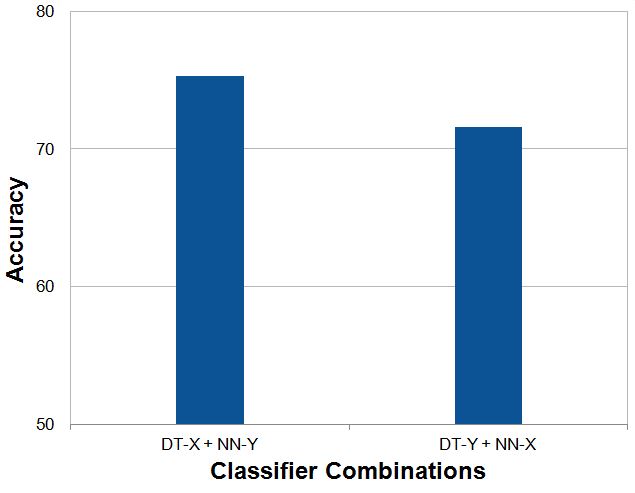}
  \caption{Mixed Classifiers in MTD where X represents the classifier with HPCs Branch-instructions, Branch-misses, Bus-cycles, and Cache-misses and Y represents the classifier with HPCs Cache-references, CPU-cycles, and Instructions.}
  \label{fig:mixclass}
\end{figure}

 In this experiment, the MTD was tested using the 50 Malware with perturbations, but the ML classifiers were mixed. We revert back to a MTD using two classifiers for this experiment. The two ML classifiers are created using different sets of HPCs. Previously, each MTD would only be run with models of the same ML algorithm. For our experiments, we modify the order in which the classifiers are built. For example, as shown in Figure 14, in the first experiment, the HPCs Branch-instructions, Branch-misses, Bus-cycles, and Cache-misses are used for the DT classifier, while the HPCs Cache-references, CPU-cycles, and Instructions are used for the NN classifier. In the second experiment, a reverse order is followed. These two classifiers are then chosen randomly using MTD.   Figure~\ref{fig:mixclass} shows the accuracy on using these two sets of classifiers.
Because the DT and NN classifiers have  accuracy values in the 70\% to 76\% range, mixing their models results in 70.9\% to 75.3\% accuracy. Thus, the detection accuracy is affected depending on how the ML classifiers are designed. The capability to mix different ML algorithms allows for an increase in the complexity of reverse engineering the MTD. However, the final accuracy of the system of mixed classifiers depends on which HPCs are used in designing a particular type of classifier.

\subsubsection{MTD Resilience}
\label{MTDresilience}
\begin{figure}[h!]
\centering
  \includegraphics[width=\linewidth]{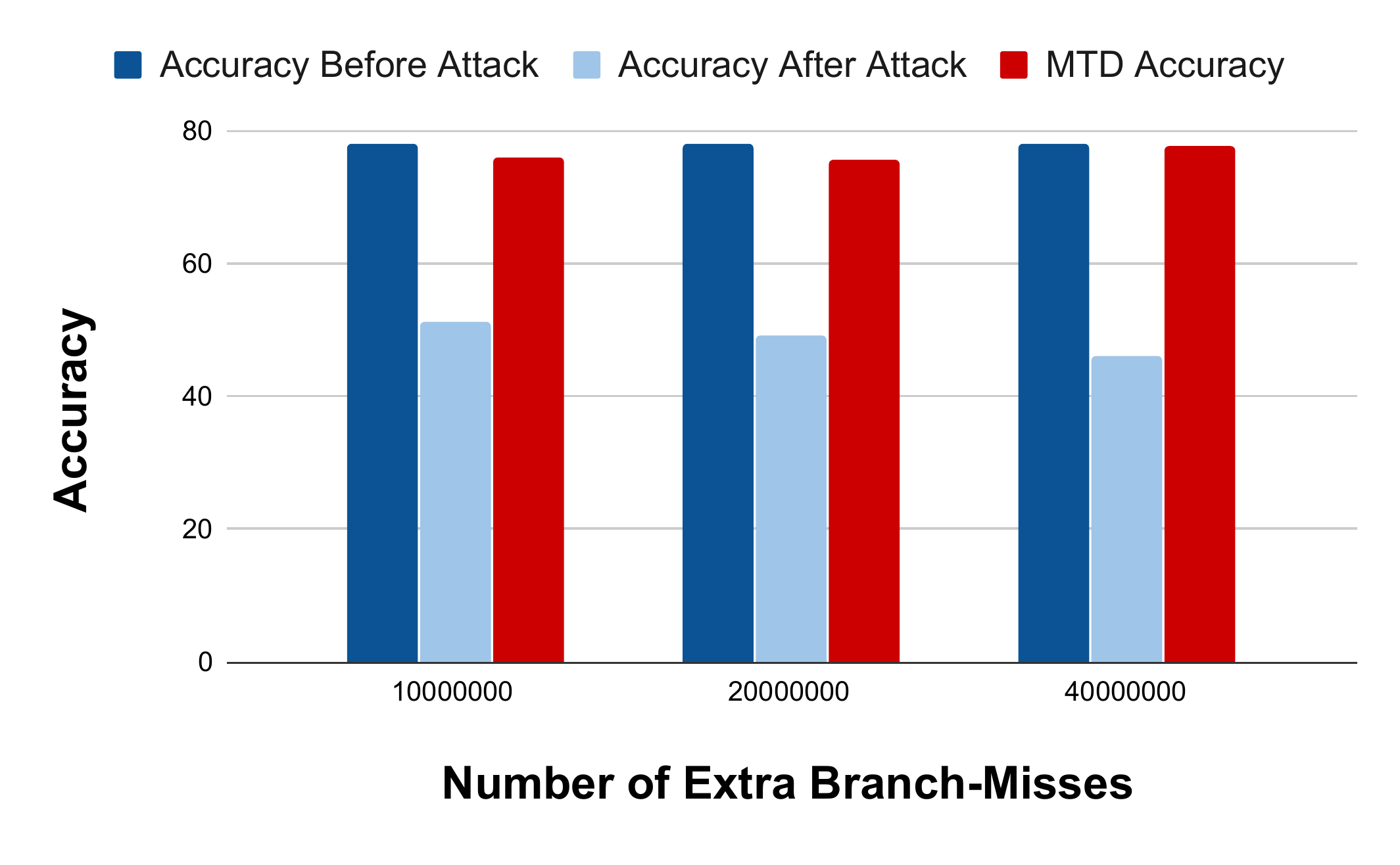}
  \caption{MTD Defense on Adversarial Attack Variants.}
  \label{fig:extrabranch}
\end{figure}

Previously, the adversarial attack reduced the accuracy of the ML classifiers to roughly 50\%. To demonstrate the resilience and durability of the MTD, 50 Malware were executed with a strengthened Adversarial Sample Generator (ASG). Figure~\ref{fig:extrabranch} shows experimental results for the adversarial attack bolstered through extra branch-misses. Initially, a trained NN classifier had a detection accuracy of 78.1\%. A set of preliminary experiments were conducted, where the ASG was set to generate branch misses less than ten million. Under such circumstances, the adversarial attack was incompetent to reduce the HMD accuracy significantly, thereby resulting in an unsuccessful attack scenario. Consequently, three sets of experiments were executed with HPC measures of branch-misses ranging from ten million to forty million, as generated by the ASG. The NN accuracy dropped from 78.1\% to 51.4\% for ten million branch-misses, 49.4\% for twenty million branch-misses, and 46.2\% for forty million branch misses. Therefore, the addition of supplementary branch-misses was successful in reducing the accuracy. However, these HPC traces were then ran through the MTD utilizing only NN classifiers where the accuracy reverted back up to 75.9\% for ten million branch-misses, 75.8\% for twenty million branch-misses, and 77.7\% for forty million branch-misses. This is equivalent to a 24.5-31.5\% increase, which is a 97.1\%-99.4\% restoration of the original accuracy. The MTD is robust against specific HPC modification adversarial attacks, because the switching of ML classifiers results in inspections of different HPCs for efficient Malware detection.   

\subsubsection{Moving Target Defense on x86}
\label{X86implementationMTD}
\textcolor{black}{
Our previous experimental data was collected from an ARM processor. To prove that our MTD works across different architectures, we have implemented the MTD on an x86 processor. We utilized 100 Malware and 100 benign programs to train base models for a Decision Tree (DT) and Neural Network (NN) classifier on Hardware Performance Counter (HPC)  branch-instructions, HPC branch-misses, HPC instructions, and HPC LLC-load-misses. Similar to Figure~\ref{fig:precaftermtd}, we used two different ML classifiers for the MTD model. One classifier was trained on HPC cpu-cycles and HPC bus-cycles. The other classifier was trained on HPC ref-cycles and HPC LLC-store-misses. We produced two different MTDs where the only difference was the ML algorithm utilized in the two classifiers. } 
\textcolor{black}{
The initial accuracy for the models was 83.6\% and 86.2\% for DT and NN. To validate the models, 15 additional Malware was tested on the DT and NN classifiers yielding an accuracy of 82.3\% and 84.3\% respectively. Consequently, these 15 Malware were executed again with the adversarial HPC generator and tested on the classifiers again. The accuracy for the DT dropped to 45.1\%, and the NN plummeted to 44.73\%. However, these same Malware with perturbations was then ran through the MTD furnishing an accuracy of 70.8\% and 72.8\%. The MTD was able to restore the DT by 25.7\% and the NN by 28.07\%. 
}

\begin{figure}[h!]
\centering
  \includegraphics[width=0.9\linewidth]{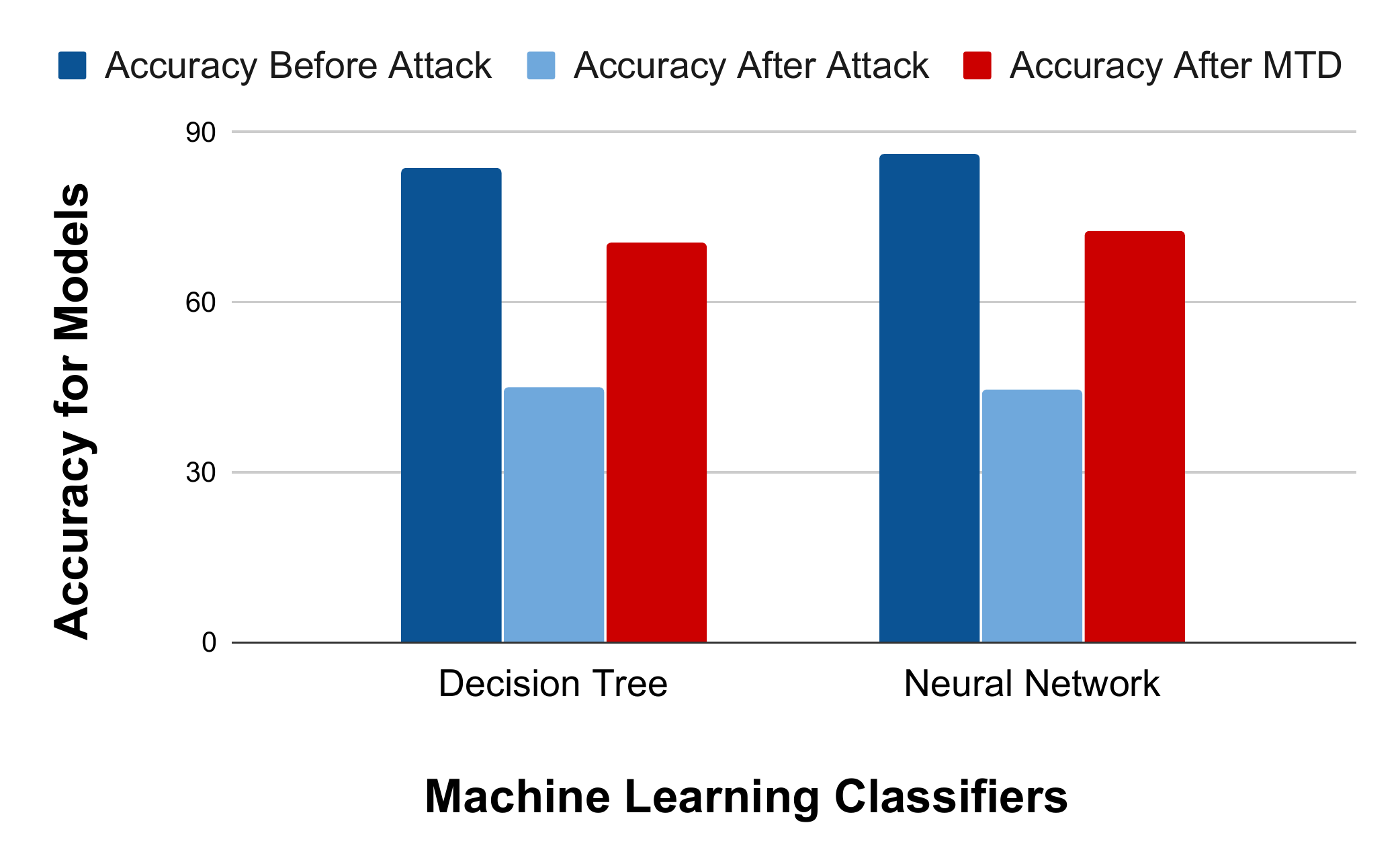}
  \caption{Accuracy with Moving Target Defense on x86.}
  \label{fig:x86accallinpaper}
\end{figure}

\subsubsection{Overhead of Moving Target Defense}
\label{overheadmetrics}
\textcolor{black}{Each MTD variant consists of two components: ML classifiers to classify an application, and a Linear Feedback Shift Register (LFSR). A LFSR is a pseudo-random number generator, on the basis of which a particular classifier is selected in each execution cycle \cite{savir1990multiple}. Since two classifiers provide optimal results, as shown in Figure~\ref{fig:classvsacc}, we implemented the hardware with two ML classifiers (DT and NN) in each case.
We designed the RTL for LFSR, Decision Trees and Neural Networks and synthesized them using the Synopsys Design Compiler logic synthesis tool. We have used two different digital standard cell libraries, lsi\_10k and saed\_90nm. The area and power overhead obtained from the synthesis of each RTL design is presented in Table~\ref{Table1ju}. Table~\ref{Table2ju} outlines the total overhead incurred to implement each MTD variant, which includes two ML classifiers (DTs or NNs) in addition to a LFSR.}

\begin{table}[h!]
\caption{Overhead of Individual MTD Components.}
\centering
\begin{adjustbox}{width=\columnwidth,center}
\begin{tabular}{|c|c|c|c|c|}
\hline
\multirow{2}{*}{\textbf{Design}} & \multicolumn{2}{c|}{\textbf{Area Overhead}}      & \multicolumn{2}{c|}{\textbf{Power Overhead}}     \\ \cline{2-5} 
                                 & \textbf{lsi\_10k} & \textbf{saed\_90nm} & \textbf{lsi\_10k} & \textbf{saed\_90nm} \\ \hline
LFSR                             & 31 sq. units      & 114.447368 $\mu m^{2}$      & 0.1406 $\mu$W      & 8.1432 $\mu$W           \\ \hline
Decision Tree 1                  & 2077 sq. units    & 8808.035814 $\mu m^{2}$     & 43.7434 $\mu$W        & 477.3366 $\mu$W         \\ \hline
Decision Tree 2                  & 2102 sq. units    & 8940.716785 $\mu m^{2}$     & 44.4456 $\mu$W        & 481.2820 $\mu$W         \\ \hline
Neural Network 1                 & 329165 sq. units  & 1620506.844 $\mu m^{2}$     & 381.2195 $\mu$W       & 36004 $\mu$W            \\ \hline
Neural Network 2                 & 330717 sq. units  & 1627408.873 $\mu m^{2}$     & 382.3155 $\mu$W       & 36021 $\mu$W            \\ \hline
\end{tabular}
\end{adjustbox}
\label{Table1ju}
\end{table}

\begin{table}[h!]
\caption{Overhead for Different MTD Variants.}
\centering
\begin{adjustbox}{width=\columnwidth,center}
\begin{tabular}{|c|c|c|c|c|}
\hline
\multirow{2}{*}{\textbf{MTD Variant}}                                                    & \multicolumn{2}{c|}{\textbf{Area Overhead}} & \multicolumn{2}{c|}{\textbf{Power Overhead}} \\ \cline{2-5} 
                                                                                         & \textbf{lsi\_10k}   & \textbf{saed\_90nm}   & \textbf{lsi\_10k}    & \textbf{saed\_90nm}   \\ \hline
\begin{tabular}[c]{@{}c@{}}Decision Tree 1 \\ + Decision Tree 2\\ + LFSR\end{tabular}    & 4210 sq. units      & 17863.19997 $\mu m^{2}$       & 88.3296 $\mu$W        & 966.7618 $\mu$W           \\ \hline
\begin{tabular}[c]{@{}c@{}}Neural Network 1 \\ + Neural Network 2 \\ + LFSR\end{tabular} & 659913 sq. units    & 3248030.164 $\mu m^{2}$       & 763.6756 $\mu$W       & 72033.1432 $\mu$W         \\ \hline
\end{tabular}
\end{adjustbox}
\label{Table2ju}
\end{table}


\section{Discussion}
\label{sec:Discussion}
We presented a MTD-based approach to counter the adversarial attack on HPC traces that misclassifies a Malware into a benign application and vice versa. The aforementioned attack ideally alters particular HPC values of a Malware or a benign program to replicate the characteristics of their respective antipodes, in order to mislead the HMD to produce an ambiguous label. In \cite{dinakarrao2019adversarial}, the adversary selects HPCs branch-misses and LLC-load-misses as target, which has dominant low values, significant of an application being a Malware. The attack then introduces dummy loops to boost up those values in order to mimic the corresponding values for a benign program, which in turn, ramps up the measures for instructions and branch instructions, deceiving the HMD.

But, Malware also tend to generate exorbitant values for some HPCs, making it impossible for the adversarial attack to reduce them altogether, since the adversarial HPC values can never be negative. \textcolor{black}{Furthermore, it is highly implausible for the adversarial sample generator to inject perturbations in more than two to three distinct sets of HPCs concurrently, due to constrained resources associated with processors deployed on edge devices.} This is where lies the major contribution of the proposed MTD model. Our proposed model takes into consideration multiple HPCs in a processor by dynamically changing the attack surface. Any inconsistencies in any of the counter measures owing to the executed application gets immediately arrested by our robust algorithm, thereby aiding the Malware detector to perform as expected. The theoretical model is corroborated by our mathematical analysis and experimental results. 
The proposed model can be generalized for any other ML classifier used in a HMD. This MTD algorithm incurs an area overhead for the ML classifiers while implementing on hardware. However, as seen from Figure~\ref{fig:classvsacc}, a total count of two classifiers provide optimal accuracy, in presence of perturbations. Therefore, in the worst case, two ML classifiers are implemented. Since this area overhead incurred is minimal compared to the processor area, as seen in Table~\ref{Table2ju}, such defense mechanism eventually will be beneficial for the CPU to protect itself from probable vulnerabilities.

\section{Conclusion}
\label{sec:Conclusion}
HMDs can improve system security by addressing the challenges of anti-virus software. In this paper, we proposed a MTD algorithm to combat HPC modification adversarial attacks in HMDs. The introduced defense strategy involves dynamic evolution of the attack surface, precisely the parameters deployed in HMD for Malware detection, thereby restoring its anticipated performance. Experimental results show that the MTD can bolster the detection accuracy by up to 31.5\%, and nearly restore the original accuracy on the HPCs that have been modified by an adversarial attack. An analytical model has been introduced proving that the probability of an attacker guessing the HPC-classifier combinations in the MTD is extremely low, near to impossible (in the range of $10^{-1864}$ for a system with 20 HPCs). In the future, we plan to 
explore utilization of different ML models to be employed in the MTD such as Naive Bayes, Logistic Regression, Support Vector Machines, and Random Forest. The effect of adversarial attacks on these algorithms and their robustness in the MTD will be analyzed.

\balance


\bibliographystyle{IEEEtran}

\begin{IEEEbiography}[{\includegraphics[width=1in,height=1.25in,clip,keepaspectratio]{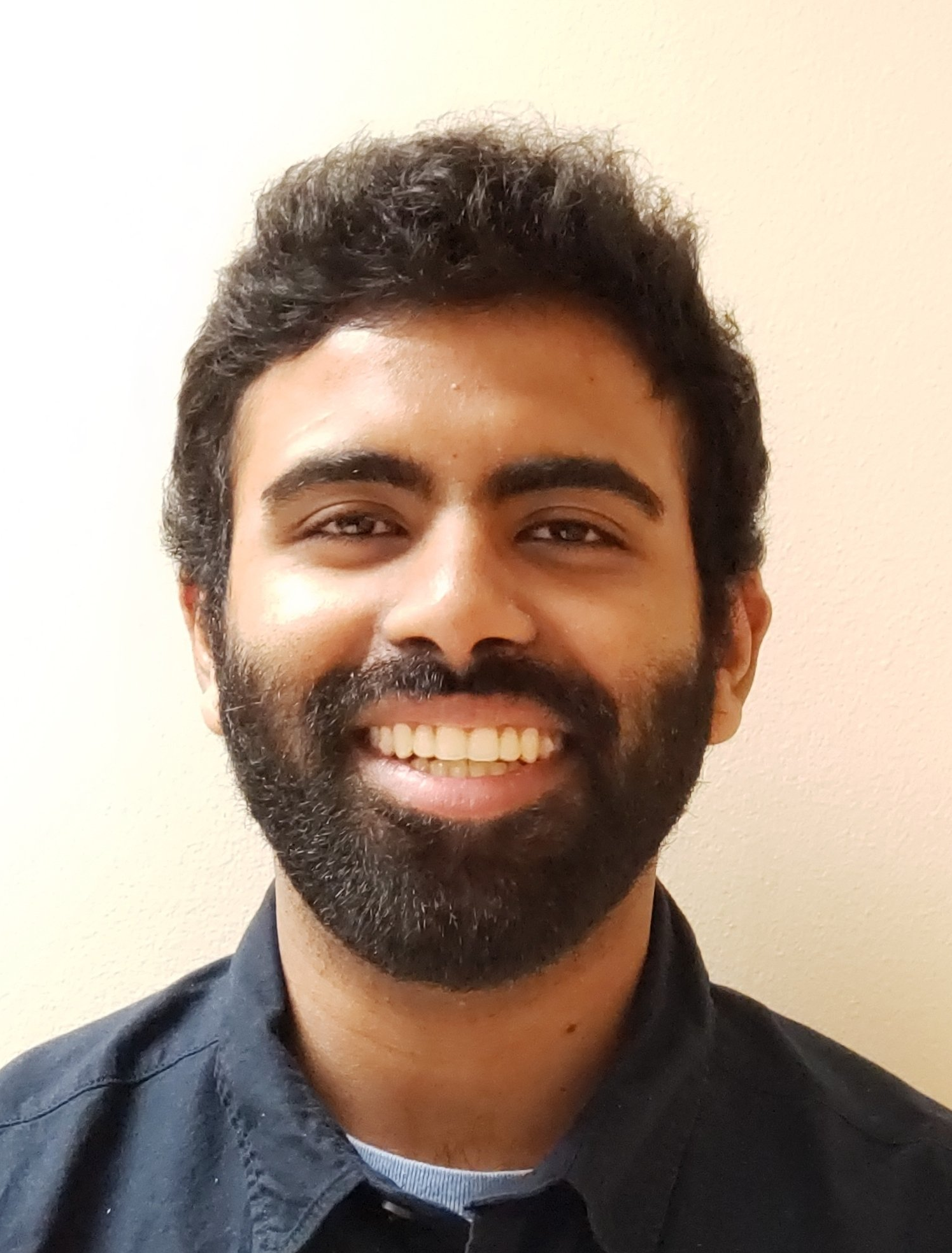}}]{Abraham Peedikayil Kuruvila} is a doctoral student in the department of Electrical and Computer Engineering at the University of Texas, Dallas. He received his BS degree in Computer Engineering from the University of California, Riverside in 2017 and MS in Computer Engineering from California State University, Fullerton in 2019. His research interests include hardware and system security, embedded systems, and hardware architectures.
\end{IEEEbiography}

\begin{IEEEbiography}[{\includegraphics[width=1in,height=1.25in,clip,keepaspectratio]{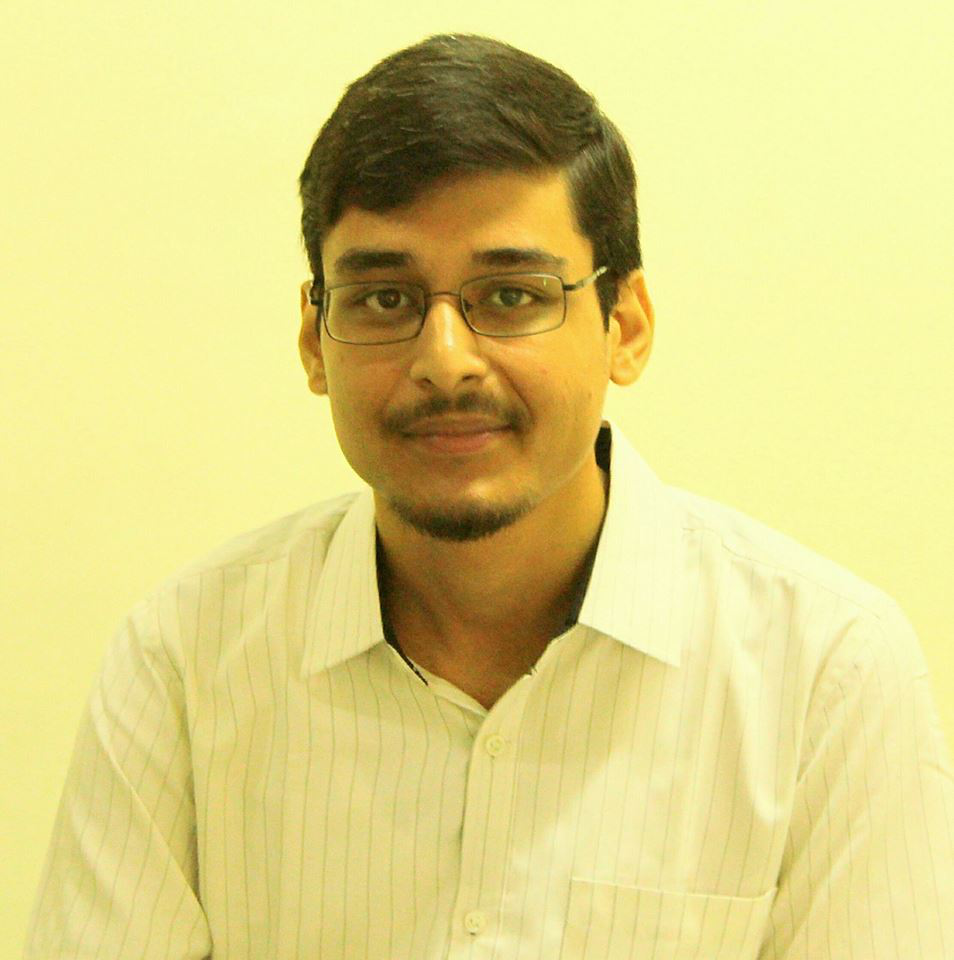}}]{Shamik Kundu} is a doctoral student in the department of Electrical and Computer Engineering at the University of Texas, Dallas. He received his B.Tech degree in Electronics and Communications Engineering from Heritage Institute of Technology in 2018. He has authored 1 journal and 2 conference articles.  His research interests include hardware and system security, fault detection and modelling in hardware architectures.
\end{IEEEbiography}

\begin{IEEEbiography}[{\includegraphics[width=1in,height=1.25in,clip,keepaspectratio]{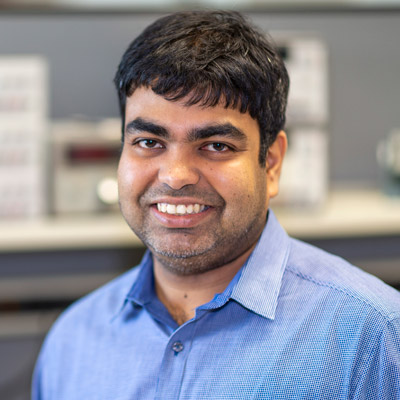}}]{Kanad Basu} received his Ph.D. from the department of Computer and Information Science and Engineering, University of Florida. His thesis was focused on improving signal observability for post-silicon validation. Post-PhD, Kanad worked in various semiconductor companies like IBM and Synopsys. During his PhD days, Kanad interned at Intel. Currently, Kanad is an Assistant Professor at the Electrical and Computer Engineering Department of the University of Texas at Dallas. Prior to this, Kanad was an Assistant Research Professor at the Electrical and Computer Engineering Department of NYU. He has authored 2 US patents, 2 book chapters and several peer reviewed journal and conference articles. Kanad was awarded the ”Best Paper Award” at the International Conference on VLSI Design 2011. Kanad’s current research interests are hardware and systems security. 
\end{IEEEbiography}

\newpage

\end{document}